\documentclass[preprintnumbers,amsmath,amssymb,floatfix,11pt,prd,onecolumn,superscriptaddress,nofootinbib]{revtex4}
\usepackage[utf8]{inputenc}
\usepackage{diagbox}
\usepackage{latexsym}
\usepackage{epsfig}
\usepackage{epstopdf,float}
\usepackage{graphicx}
\usepackage{amssymb}
\usepackage{amsmath}
\usepackage{dcolumn}
\usepackage{bm}
\usepackage{color}
\usepackage{comment}
\usepackage[shortlabels]{enumitem}
\usepackage{subfigure}
\usepackage{multirow}
\usepackage{xcolor}
\begin{document}
\title{\bf Probing $f(R)$ AdS Black Hole via Hawking Evaporation, Shadows and Thermal Fluctuations}
\author{Muhammad Israr Aslam}
\altaffiliation{mrisraraslam@gmail.com,
israr.aslam@umt.edu.pk}\affiliation{Department of Mathematics,
School of Science, University of Management and Technology,
Lahore-$54770$, Pakistan.}\author{Saira Waheed}
\altaffiliation{swaheed@pmu.edu.sa}\affiliation{Prince Mohammad Bin
Fahd University, Al Khobar, Kingdom of Saudi Arabia}\author{Rabia
Saleem} \altaffiliation{rabiasaleem@cuilahore.edu.pk}
\affiliation{Department of Mathematics, COMSATS University
Islamabad, Lahore Campus, Lahore-54000, Pakistan}
\author{Nazek Alessa} \altaffiliation{naalessa@pnu.edu.sa}\affiliation{Department
of Mathematical Sciences, College of Science, Princess Nourah bint
Abdulrahman University, P.O.Box 84428, Riyadh 11671, Saudi Arabia}

\begin{abstract}
The process of Hawking evaporation, shadows and thermal fluctuations
are investigated within the fabric of $f(R)$ AdS Black Hole (BH).
Specifically, the Hawking evaporation process is analyzed
numerically using the Stefan-Boltzmann law. The results indicate
that the BH lifetime is always infinite, which means the BH becomes
a remnant in the late time. Additionally, the evaporation rate
depends on the AdS radius $\ell$ and coupling parameters, both
$\lambda$ and $\psi_{0}$. We further examined the visual properties
of BH shadows observed for various values of both $\lambda$ and
$\psi_{0}$. The results reveal that the BH shadow radius decreases
with $\psi_{0}$, while it increases with $\lambda$. Consequently,
we further investigate the infalling accretion matter in the
vicinity of BHs. The results depict that while variations in
relevant parameters do influence the central region, the important
factor is the change in optical appearance of the bright photon
ring, which is exhibited at the position of the photon sphere. Next, we
discuss many thermodynamical quantities, such as temperature,
entropy, Helmholtz free energy, internal energy, corrected pressure,
enthalpy, Gibbs free energy and specific heat and interpret how
the variations in $\lambda$ and $\psi_{0}$
impact on the stability and phase transitions of the AdS BHs.\\

\end{abstract}

\date{\today}

\maketitle

\section{Introduction}

In the fabric of quantum field theory for a curved spacetime,
Hawking confirmed that the effects arising from the quantum
mechanics enable BHs to radiate particles, today's so-called
phenomena of Hawking radiations \cite{ep1,ep2}. The root cause
behind this radiation is the quantum mechanics, which assures a
continuous creation of virtual pairs of particles and anti-particles
closer to the black hole event horizon. It is argued that typically
vacuum state is easily achieved as these pairs of particles are
annihilated by each other; however, Hawking radiation emerges when
one pair member is trapped by the black hole while its counterpart
escapes and hence results in loss of BH mass and energy. Hawking
radiation together describes the dynamics of quantum physics and
gravity, which constitutes a realistic path to recognize the further
significant attributes of quantum gravity. The continuous emission
of radiation from the interior of a BH, leading to the loss of mass and
thermal entropy, shrinking its event horizon and increasing its
surface gravity, which further leaves us with the problematic
information loss paradox \cite{ep3,ep4}. The phenomena of Hawking
radiation facilitate the calculation of the emission power of
particles and to approximate the lifetime of the BH. It is a
well-known fact that a BH with a larger mass corresponds to a slower
evaporation rate, and thus Hawking radiation becomes particularly
noteworthy in the case of small BHs.

On theoretical landscape, the phenomenon of Hawking evaporation has
revolutionized our current understanding of the cosmos, as it implies
that BH being a dynamical object plays a worthwhile role, and this
process may help to explain a wide range of phenomena in physics, like BH thermodynamics and the information paradox. Significant studies
are available in the literature \cite{SN1,Robinson} revealing the impact
of Hawking radiation in the early cosmos and how the presence of
gravitational anomalies affects the Hawking evaporation process. In
this regard, one can calculate the Hawking emission rate of the BH,
with the help of Stefan Boltzman law \cite{ep5}. In the framework of
Einstein gravity, the lifetime of a static BH has $t\sim M_{0}^{3}$,
which is associated with the initial mass $M_{0}$ of the BH
\cite{ep6}. From this relation, it follows that if a black hole has
an infinite initial mass, its lifetime diverges. Similar approach
can also be considered for rotating \cite{ep7} or charged BHs
\cite{ep5,ep8}. The rotating BH loses angular momentum much faster
than its mass and transforms it into a static Schwarzschild BH
(unless there are a lot of scalar particle species \cite{ep9,ep10}).
Whereas the charged BH loses charge as well as mass, but the
charge-to-mass ratio can grow, tending toward an extreme black hole.
In this scenario, the lifetime of these BHs may extend
significantly, but finally transform into the Schwarzschild state
\cite{ep5}. In fact, the process of Hawking evaporation provides a
realistic way to understand the essential properties of BH. In
\cite{ep11}, authors studied the BH behavior with the help of
quantum effects numerically and concluded that the Hawking radiation
can be extended in certain areas. de Rham and Zhang \cite{ep12}
explored the effects of rotation on the Hawking radiation of BHs and
concluded that it can lead to interesting new phenomena, such as the
emission of gravitational waves. Considering the Bardeen space-time
as quantum corrections to the Schwarzschild-like BH solution, the
authors in \cite{ep13} explored the quasi-normal modes and Hawking
radiation and concluded that the intensity of Hawking radiation is
substantially reduced by up to three orders of magnitude due to
quantum corrections. Consequently, in recent studies, various
significant properties of Hawking evaporation of AdS BHs are
explored in the context of modified theories of gravity, for
instance, one can see Refs.
\cite{massive,stvg,conformal,ong,HeGB,Jawad,lifshitz}.

Particle movement within a system can give rise to arbitrary
temperature or energy fluctuations, which are known as thermal
fluctuations, playing a noteworthy role in various phenomena such as
BH physics, especially in Hawking evaporation of BHs. It is argued
that the production of particle pairs closer to the event horizon may be
caused by these energy and temperature fluctuations. It is also
pointed out that in contemporary astrophysics, these thermal
fluctuations have a strong impact on the characteristics of emitted
radiation, BHs evolution, as well as their distribution in our
cosmos. The fact that BHs possess temperature validated Bekenstein's
idea about BH entropy, which is further linked to the event horizon
area \cite{SN2,SN3} through the relationship $S=\frac{A}{4}$. In
this respect, quantum fluctuations resulted in the corrections to
the maximum BH entropy, which opened doors for the formation of
the holographic principle \cite{SN4}. This principle states that the
information contained inside a spatial domain can be represented on
its boundary, with the density limited to one degree of freedom per
Planck cell at maximum \cite{SN5*}. To this end, numerous strategies
have been implemented to evaluate entropy corrections, e.g.,
non-perturbative quantum GR. In the literature, a lot of work has been
done on this subject.

Jing and Yan computed the entropy corrections for dilation BHs and
concluded that these corrections are of logarithmic type \cite{SN5}.
BH global monopole structures exhibiting static and spherical
symmetry originated from the Dirac spinor field has been examined by
Maowang and Ji \cite{SN6} through the brick wall method and
interesting outcomes have been achieved. In another work, Gour and
Medved \cite{SN7} explored the impact of thermal effects on entropy
of charged as well as uncharged BHs. Several recent studies
\cite{SN8,SN8o1,SN8o2,SN9,SN9o1} have provided new insights into
entropy corrections arising from thermal or quantum fluctuations in
various black hole models, including Schwarzschild black holes, BTZ
black holes in asymptotically AdS and dS spacetimes, charged AdS
black holes, and Hayward black holes. In this context, Ji
\cite{SN10} studied the rotating BH and its emitted gravitational
radiation under thermal fluctuations and found that such
fluctuations can cause gravitational radiation to increase and
also, their impact becomes more obvious on rapidly spinning BHs. The
link between BH entropy and thermal fluctuations has been
investigated by Ma et al. \cite{SN11} and they found that entropy
fluctuations of BHs arising from thermal effects are directly
related to the square of the black hole temperature. In the
surrounding of the BH event horizon, authors \cite{SN12} investigated
the characteristics of thermal fluctuations which may give rise to
the generation of entangled particle pairs plays a significant
role in comprehending BH physics. In Born-Infeld massive gravity,
Jawad \cite{SN13} studied the thermal stability of two BHs under
thermal fluctuations: AdS BH with non-abelian hair and AdS BH with
charge as well as a global monopole. The author examined various
significant thermodynamic measures graphically for these BHs and
discussed phase transitions, local as well as global stability of
these BHs.

The existence of BHs in our universe has long been one of the most
intriguing topics in modern cosmology. These massive objects were
predicted theoretically as solutions of the Einstein field
equations, but definitive observational confirmation was absent.
Recently, researchers have found strong evidence about the existence
of BHs through gravitational signals detected by LIGO \cite{SN14}
and later, event horizon telescope (EHT) captured the image of BH
(possibly located at the center of the Milky Way Galaxy M87) as
shadow surrounded by a bright ring-shaped radiation lump
\cite{SN15}. This remarkable discovery paved the way for deeper
exploration of BH physics within the research community. Since a
For a long time, researchers have been exploring the subject of WH shadows
(mainly caused by light deflection because of strong gravity)
and its related concepts in various modified gravity theories. In
this respect, Narayan et al. \cite{SN16} examined the Schwarzschild
BH's shadow and concluded that the observed optical properties of
an accreting BH are unaffected by the radius of the inner region at
which radiation from the gas ends. Gralla et al. \cite{SN17}
discussed the visual signature of radiating gas closer to the BH casting
a shadow surrounded by a bright photon ring. Israr and Rabia
\cite{SN18} explored the shadow cast by a dyonic BH in the
presence of a global monopole along with a perfect fluid by considering
different accretion profiles. In another work \cite{SN19}, authors
studied Einstein images produced by charged Rastall AdS BH through
the framework of holography. In an interesting recent study, Zeng et
al. \cite{SN21} discussed the shadows cast by a Kerr-like BH in
the presence of a cold dark matter halo subject to illumination from a
celestial source along with a thin accretion disk. Additionally,
various studies have focused on the influence of thin and thick
accretion disk on the optical appearance of BH shadows
\cite{sd34,gkref26,gkref27,thin1,thick1,thick2}, including the wave
optics mechanism
\cite{sd22,israr1,israr22,prog,frgravity,He2024PDU,SN20}.

To comprehend the speedy cosmic expansion, the well-established idea
is to incorporate some additional degrees of freedom representing
the dark ingredients in the gravitational segment of GR action and
the resulting frameworks are labeled as modified gravity theories.
Recently, a bulk of captivating modified theories have been put
forward by researchers, which are further tested on different
cosmological grounds and proved to be quite successful. In this
respect, the most genuine extension of GR is the well-known $f(R)$
gravity, which has effectively resolved many outstanding cosmic
problems. Efforts to obtain BH solutions persist in the context of
$f(R)$ theory, leading to the proposal of some compelling models
\cite{SN22,SN23,SN24,action,ma1,ma2}. Being inspired by the above
highlighted literature, the current work aims to study Hawking
evaporation, shadows and thermodynamic quantities of $f(R)$ AdS BH.
Upcoming segments are sequenced in the following pattern. Section
\textbf{II} defines AdS $f(R)$ BH as well as some
requirements needed to accomplish the task. In Section \textbf{III},
we shall explain the Hawking evaporation process of the considered
BH graphically. The next section will give a graphical analysis of
unstable photon orbits and observed intensities through radiating
gas. Section \textbf{V} computes some noteworthy thermal quantities
for AdS $f(R)$ BH and examines their graphical behavior. The final
segment concludes the whole study and illuminates our main findings.

\section{AdS Black Hole in $f(R)$ Theory with a Global Monopole}

In the $f(R)$ theory of gravity, the action that shows an interaction with a matter field is defined as \cite{action}
\begin{equation}\label{1}
S=\frac{1}{2\kappa}\int\sqrt{-g}d^4x (f(R)+\mathcal{L}_{m}),
\end{equation}
where $f(R)$ is an analytic function of Ricci scalar $R$, $k=8\pi G$ ($G$ denotes
the gravitational constant), $g$ is the determinant of the metric
tensor and $\mathcal{L}_{m}$ is the ordinary matter Lagrangian density. Through the metric
variation of the action $S$, the field equations can be derived as
\begin{eqnarray}\label{2}
k T_{\xi\zeta}&=&F(R)R_{\xi\zeta}-\frac{1}{2}f(R)g_{\xi\zeta}-(\nabla_{\xi}\nabla_{\zeta}-g_{\xi\zeta}\Box)F(R),
\end{eqnarray}
where $F(R)=df(R)/dR$,~$\Box=\nabla_{\mu}\nabla^{\mu}$ and
$T_{\xi\zeta}$ is the energy-momentum tensor. The Lagrangian density
with the coupling of the global monopole can be defined as
\begin{eqnarray}\label{3}
\mathcal{L}_{m}=\frac{1}{2}\partial_{\xi}\phi^{\alpha}\partial^{\xi}\phi^{\alpha}-
\frac{1}{4}\eta(\phi^{\alpha}\phi^{\alpha}-\lambda^{2})^{2},
\end{eqnarray}
in which $\eta$ is the coupling constant that vanishes in the
energy-momentum tensor \cite{ma1}, $\lambda$ denotes the spontaneous
breaking of $\mathcal{O}(3)$ symmetry to $U(1)$ symmetry and
$\phi^{\alpha}$ is given by an isotriplet of scalar fields
corresponding to the well-known hedgehog ansatz \cite{ma1}. Now
considering the above field equations and using some appropriate
approximations with the addition of the cosmological constant
$\Lambda$ (for details see Refs \cite{action,ma1,ma2}), one can
obtain a static AdS BH solution in $f(R)$ gravity as given below:
\begin{eqnarray}\label{4}
ds^{2}=-A(r)dt^{2}+\frac{dr^{2}}{A(r)}+r^{2}d\theta^{2}+r^{2}\sin^{2}\theta
d\varphi^{2},
\end{eqnarray}
with
\begin{equation}\label{5}
A(r)=1-\frac{2M}{r}+3M\psi_{0}-\psi_{0}
r-8\pi\lambda^{2}+\frac{\Lambda}{3}r^{2},
\end{equation}
where $M$ refers to the mass of BH, parameter $\psi_{0}$ originates from
the $f(R)$ gravitational framework and $\psi_{0} r$ illustrates the
deviation from GR. In GR, we define $F(R)=df(R)/dR$ and $F(R)=1$.
Moreover, $R$ is a function of $r$, consequently, it can be written as
$F(R)=F(R(r))=\mathcal{F}(r)$. Now, by adopting an alternative
parametrization for $F(R)$, we have $\mathcal{F}(r)=1+\psi$, where
$\psi=\psi_{0} r$. In addition, the cosmological constant
$\Lambda=3/\ell^{2}$, where $\ell$ is the length of AdS radius. By
setting $A(r=r_{+})=0$, one can obtain the mass of BH as
\begin{equation}\label{6}
M=\frac{r_{+}(\ell^{2}+r_{+}^{2}-8\pi\ell^{2}\lambda^{2}-\ell^{2}r_{+}\lambda)}{\ell^{2}(2-3r_{+}\psi_{0})},
\end{equation}
where $r_{+}$ is the event horizon of the BH. The temperature of the
BH is proportional to the surface gravity at the event horizon,
which is calculated as
\begin{eqnarray}\label{7}
T=\frac{1}{4\pi}\frac{dA}{dr}\Big|_{r=r_{+}}=\frac{6r^{2}_{+}(\psi_{0}r_{+}-1)+\ell^{2}
(16\pi\lambda^{2}-2+\psi_{0}r_{+}(4-3\psi_{0}r_{+}))}{4\ell^{2}\pi
r_{+}(3\psi_{0}r_{+}-2)},
\end{eqnarray}
which clearly depends on the AdS radius $\ell$, $f(R)$ parameter $\psi_{0}$
and the global monopole parameter $\lambda$.
\begin{figure}[H]\centering
\subfigure[\tiny][]{\label{t1}\includegraphics[width=5.4cm,height=4.5cm]{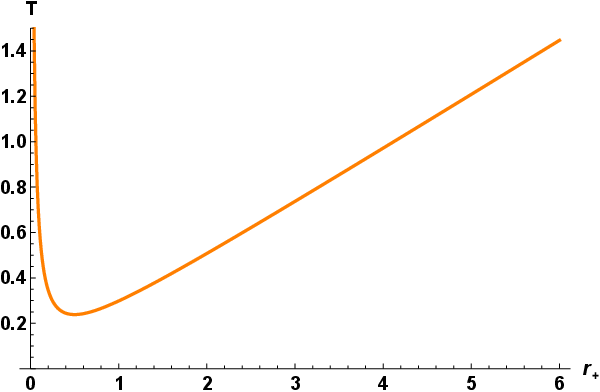}}
\subfigure[\tiny][]{\label{t2}\includegraphics[width=5.4cm,height=4.5cm]{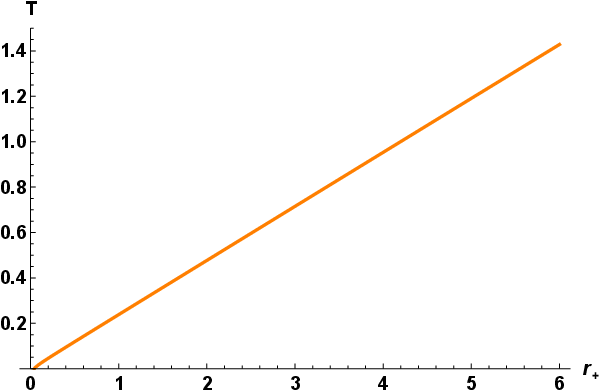}}
\subfigure[\tiny][]{\label{t3}\includegraphics[width=5.4cm,height=4.5cm]{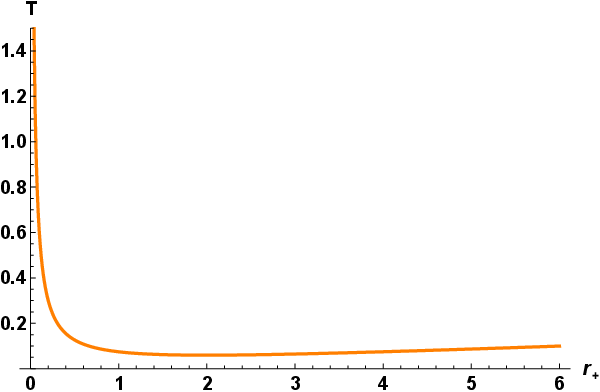}}
\caption{Plot $T$ vs $r_{+}$ corresponding to
$\psi_{0}=0.001$,~$\lambda=0.1$,~$\ell=1$ (a),
$\psi_{0}=-0.001$,~$\lambda=0.2$,~$\ell=1$ (b), and
$\psi_{0}=0.001$,~$\lambda=0.1$,~$\ell=4$ (c).}\label{evp1}
\end{figure}
In Fig. \textbf{\ref{evp1}} we depict the behavior of temperature
$T$ with respect to the horizon radius $r_{+}$ for some particular
choices of parameters as an examples. In Fig. \textbf{\ref{evp1}}
(a), we see that initially the temperature $T$ has the maximum value
but suddenly decreases at $r_{+}\approx0.44$ leads that
$T\approx0.2404$ and then sharply increases with the aid of the
$r_{+}$. The development of temperature $T$ as shown in Fig.
\textbf{\ref{evp1}} (b) is quite different from the previous one. As
the temperature $T$ starts from zero at $r_{+}=0$ and goes to
infinity when $r_{+}\rightarrow\infty$. Further for large AdS, such
as when $\psi_{0}=0.001$,~$\lambda=0.1$,~$\ell=4$, we see that
initially the temperature $T$ has a maximum value and then decreases
and behaves almost constantly with respect to $r_{+}$, see Fig.
\textbf{\ref{evp1}} (c). The Bekenstein-Hawking entropy, which can
be evaluated by using the first law of BH thermodynamics $dS=dM/T$,
provides the standard area law $S=\pi r_{+}^{2}$.

\section{Hawking Evaporation of $f(R)$ AdS Black Hole}

After the analysis of thermodynamic quantities of $f(R)$ AdS BH with global monopole, we
are going to discuss the Hawking evaporation of BH. As it is a well-known fact that, due to
the emissions of Hawking radiation, the BH mass declines with respect to time $t$. In this
regard, it is considered that all the emitted particles move along the null geodesics, therefore
one can apply the geometric optics approximation. In general, we consider that the motion of
massless particles located at the equatorial plane such that $\theta=\pi/2$. So, the Lagrangian
$\mathcal{L}$ for a particle in this space-time can be defined as
\begin{eqnarray}\label{8}
\mathcal{L}=\frac{1}{2}g_{\mu\nu}\dot{x}^{\mu}\dot{x}^{\nu},
\end{eqnarray}
herein $\dot{x}^{\mu}$ is the four velocity of a particle and the ``dot'' refers to differentiation of
the quantity with respect to the affine parameter $\tau$. One can define the conjugate momenta
$\rho_{\mu}=\frac{\partial \mathcal{L}}{\partial \dot{x}^{\mu}}$, which, further, provides two
conserved quantities such as the energy $E$ and angular momentum $L$ of the massless particle as follows
\begin{eqnarray}\label{9}
\rho_{t}=E=A(r)\dot{t},\quad \rho_{\varphi}=L=r^{2}\dot{\varphi}.
\end{eqnarray}
With the help of these two quantities, we define the impact parameter as $b=L/E$. Now, we have the geodesic
equation of the massless particles as \cite{ep5}
\begin{equation}\label{10}
\dot{r}^{2}=\frac{1}{b^{2}}-\frac{A(r)}{r^{2}}.
\end{equation}
Consider that the distant observer, located on the AdS boundary, is unable to observe a particle radiated
in the neighbourhood of the BH horizon due to the presence of a turning point, which satisfies $\dot{r}=0$ and
$\ddot{r}=0$. Further, the massless particle approaches to infinity when $\frac{1}{b^{2}}>\frac{A(r)}{r^{2}}$,
for all $r>r_{+}$ and the critical impact parameter $b_{c}$ can be determined at the peak position of
the $\frac{A(r)}{r^{2}}$. After calculating the expression of the $b_{c}$, one can evaluate the Hawking
evaporation of the BH with the help of the Stefan-Boltzmann law, and it can be written as
\begin{equation}\label{11}
\frac{dM}{dt}=-\hat{\sigma}\mathcal{G}b_{c}^{2}T^{4},
\end{equation}
where $\hat{\sigma}=\frac{\pi^{3}k^{4}}{15c^{3}\hbar^{3}}$. Since we are only interested in talking about
the qualitative aspects of the evaporation mechanism, so we will absorb this factor into grey-body factor
$\mathcal{G}$ and hence, we set $\hat{\sigma}\mathcal{G}=1$. Now, we are in a position to analyze the BH
evaporation procedure for different features of $T$ and $b_{c}$. In this scenario, the emission power
in $4$-dimensional space-time is proportional to the $2$-dimensional cross section $b_{c}^{2}$ and the
energy density of the photon $T^{4}$ lies in $3$-dimensional space. Significantly, the asymptotic response
of temperature $T$ has essential significance in the BH evaporation process due to the highest order of $T^{4}$.
Introducing the dimensionless variables \footnote{Of course, in our units, everything
is dimensionless, depends on the one may choose.} such as, $x=\frac{r_{+}}{\ell}$,~ $y=\psi_{0}\ell$ and
$\lambda^{\star}=\lambda$ to obtain $M\sim l$,~$T\sim l^{-1}$,~$b_{c}\sim l$ and $\ell\sim l$, in which $l$
is some dimension length, the mass $M$ and temperature $T$ of BH can be, respectively, obtained in
dimensionless variables form as follows
\begin{eqnarray}\label{12}
M&=&\frac{\ell x(x(y-x)+8\pi\lambda{^{\star}}^{2}-1)}{3xy-2},
\\\label{13}
T&=&\frac{3x^{2}y^{2}-6x^{3}y-4xy+6x^{2}-16\pi\lambda{^{\star}}^{2}+2}{4\pi\ell
x(2-3xy)}.
\end{eqnarray}
In order to obtain the expression of $b_{c}$, one can solve $\frac{\partial}{\partial r}\frac{A(r)}{r^{2}}=0$,
which produces two roots as given below
\begin{eqnarray}\nonumber
r_{p1}&=&\frac{2-16\pi\lambda^{2}+6M\psi_{0}-\sqrt{(16\pi\lambda^{2}-6M\psi_{0}-2)^{2}-24M\psi_{0}}}{2\psi_{0}},\\\label{14}
r_{p2}&=&\frac{2-16\pi\lambda^{2}+6M\psi_{0}+\sqrt{(16\pi\lambda^{2}-6M\psi_{0}-2)^{2}-24M\psi_{0}}}{2\psi_{0}},
\end{eqnarray}
where $r_{p2}>r_{p1}$ and we use $r_{p2}$ for more comprehensive analysis. This radius is associated
with the peak value of the effective potential $V(r)=\frac{A(r)}{r^{2}}$. An interesting observation
is that the critical orbits remain unaffected by the AdS radius or cosmological constant. Now the
critical impact parameter $b_{c}=\frac{r_{p2}}{\sqrt{A(r_{p2})}}$ can be defined in terms of dimensionless
variables as
\begin{eqnarray}\nonumber
b_{c}&=&\bigg(4\bigg(-\frac{y^{2}}{2}\sqrt{f_{1}+f_{2}}-\frac{4xy^{3}(x(y-x)-1+
8\pi\lambda{^{\star}}^{2})}{(3xy-2)(2+\sqrt{f_{1}+f_{2}}+f_{3}-
16\pi\lambda{^{\star}}^{2})}+\bigg(1-
8\pi\lambda{^{\star}}^{2}+\frac{\sqrt{f_{1}+f_{2}}}{2}\\\label{15}&+&\frac{3xy(x(y-x)-1+
8\pi\lambda{^{\star}}^{2})}{3xy-2}\bigg)^{2}\bigg)
\bigg(\bigg(2+\sqrt{f_{1}+f_{2}}+f_{3}-16\pi\lambda{^{\star}}^{2}\bigg)^{2}\bigg)^{-1}\bigg)^{-\frac{1}{2}}\ell,
\end{eqnarray}
where
\begin{eqnarray}\nonumber
f_{1}&=&\frac{4\big(2+3x^{2}y(x-y)-16\pi\lambda{^{\star}}^{2}\big)^{2}}{(2-3xy)^{2}},\\\nonumber
f_{2}&=&\frac{24xy(1+x^{2}-xy-8\pi\lambda{^{\star}}^{2})}{3xy-2},\\\nonumber
f_{3}&=&\frac{6xy(x(y-x)-1+8\pi\lambda{^{\star}}^{2})}{3xy-2}.
\end{eqnarray}
Inserting the expressions of $M$,~$T$ and $b_{c}$ in Eq. (\ref{11}), we derive the relationship
governing the lifetime of BH evaporation as
\begin{equation}\label{16}
dt=\ell^{3}B(x,y,\lambda^{\star})dx,
\end{equation}
where $B(x,y,\lambda^{\star})=-\frac{\partial
M(x,y,\lambda^{\star})}{\partial x}\frac{1}{b_{c}^{2}(x,y,\lambda^{\star})T^{4}(x,y,\lambda^{\star})}$ is
a complicated function, which need not be stated explicitly. We numerically solve the above expression through
Wolfram Mathematica to obtain the desired results. If we set $y$ to a constant and integrate the equation
from $\infty$ to $x_{min}=0$, the resulting BH lifetime is approximately of order $\sim\ell^{3}$.
\begin{figure}[H]\centering
\subfigure[\tiny][]{\label{t1}\includegraphics[width=5.35cm,height=5.9cm]{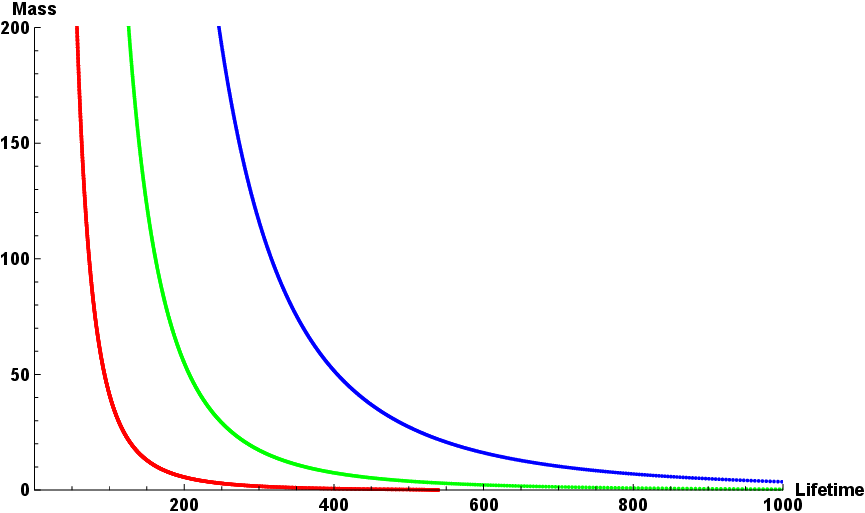}}\hfil
\subfigure[\tiny][]{\label{t2}\includegraphics[width=5.35cm,height=5.9cm]{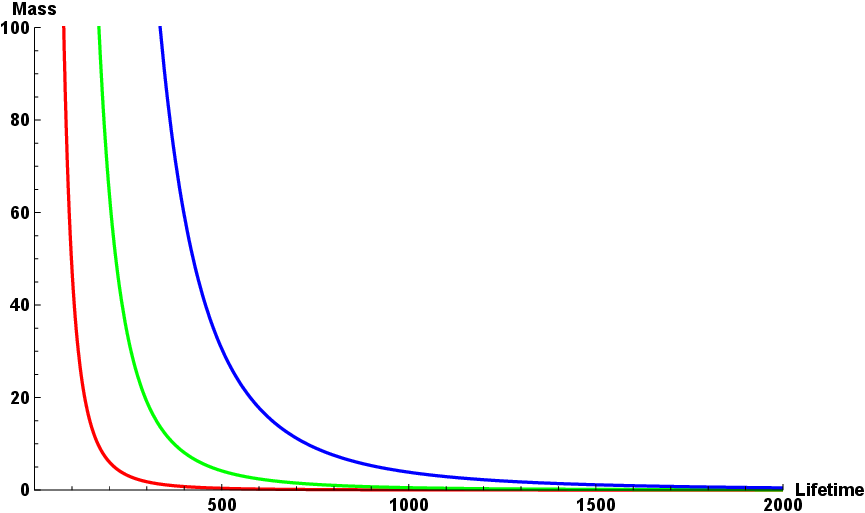}}
\subfigure[\tiny][]{\label{t2}\includegraphics[width=5.35cm,height=5.9cm]{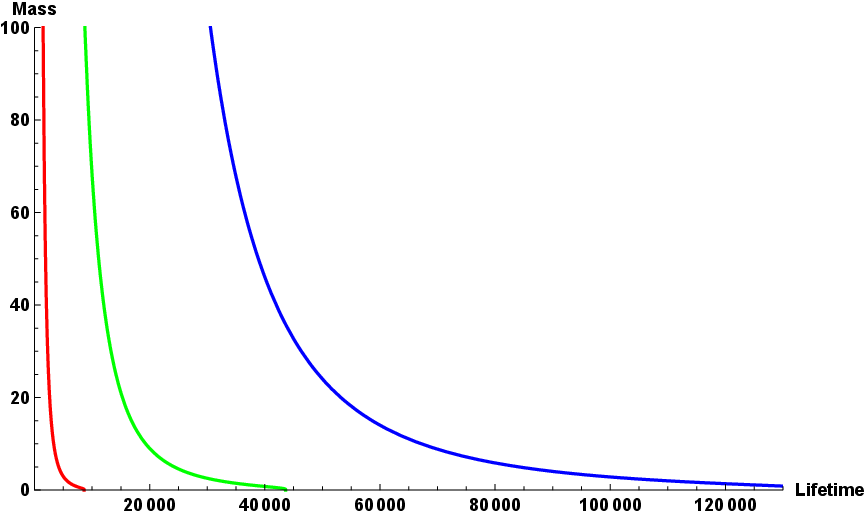}}
\caption{The numerical evaluation of the $f(R)$ AdS BH mass $M$ as a
function of the lifetime $t$ with $y>0$ and $y<0$ for figure (a) and
(b), respectively. In figure (a), we fix
$y=0.001$,~$\lambda^{\star}=0.1$ with $\ell=1\rightarrow$
\textcolor{red}{$\spadesuit$},~$\ell=1.2\rightarrow$
\textcolor{green}{$\spadesuit$} and $\ell=1.4\rightarrow$
\textcolor{blue}{$\spadesuit$}. In figure (b), we fix
$y=-0.001$,~$\lambda^{\star}=0.2$ with $\ell=1\rightarrow$
\textcolor{red}{$\spadesuit$},~$\ell=1.2\rightarrow$
\textcolor{green}{$\spadesuit$},~ $\ell=1.4\rightarrow$
\textcolor{blue}{$\spadesuit$} and in the figure (c), we fix
$y=0.001$,~$\lambda^{\star}=0.1$ with $\ell=2\rightarrow$
\textcolor{red}{$\spadesuit$},~$\ell=3\rightarrow$
\textcolor{green}{$\spadesuit$} and $\ell=4\rightarrow$
\textcolor{blue}{$\spadesuit$}.}\label{evp2}
\end{figure}
Now we graphically examine the process of Hawking evaporation for different sets of model parameters,
such that the significant features can be nicely plotted. Sometimes the values of model parameters are too
small, i.e., $M<1$ is less than a Planck mass in our units, and there is no other reason to expect that
BH evaporation still satisfy the usual Stefan-Boltzmann law at the Planck scale. Our choices of model
parameters only for convenience, there is no other physical meanings. Previously, as depicted in
Fig. \textbf{\ref{evp1}}, we discussed the physical interpretation of the BH temperature $T$, which
plays an important role in BH evaporation due to its highest-order contribution. The term $T=0$ yields
the $r_{min}$ radius which help us to find the critical mass of the BH such as $M_{min}=M(r_{min})$,
and BH evaporates within a finite timescale, i.e., changes the mass $M$ to $M_{min}$. In Fig. \textbf{\ref{evp2}},
we numerically present some examples of the BH evolution. Particularly, when $y=0.001$,~$\lambda^{\star}=0.1$
and from left to right curves corresponds to $\ell=1$,~$\ell=1.2$ and $\ell=1.4$, respectively, it can be
seen that the mass of the BH decreases rapidly at early times; however, evaporation becomes increasingly
suppressed as the BH mass and temperature decrease, approaching $T\rightarrow0$, thus the BH will
take infinite time to completely evaporate away, see Fig. \textbf{\ref{evp2}} (a). This phenomenon also
satisfies the third law of BH thermodynamics, i.e., the BH becomes a remnant, may assisting us in resolving
the information paradox \cite{ma3}. The quite similar results are also depicted in Fig. \textbf{\ref{evp2}} (b),
where the lifetime of BH is larger as compared to previous one. In this case, the values of AdS radius $\ell$
are same from left to right curves however, we fix $y=-0.001$,~$\lambda^{\star}=0.2$ for feasible plotting.
On the other hand, when we fix $y=0.001$,~$\lambda^{\star}=0.1$ and from left to right curves corresponds to
$\ell=2$,~$\ell=3$ and $\ell=4$, respectively, the BH can always evaporate away in a finite amount of time,
as shown in Fig. \textbf{\ref{evp2}} (c). This phenomena is quite different from Figs. \textbf{\ref{evp2}} (a)
and \textbf{\ref{evp2}} (b), as the lifetime of BH is huge for larger AdS.

\section{Analysis of the Unstable Photon Orbits and Observed Intensities Via Radiating/Infalling Gas}

\begin{figure}[H]\centering
\subfigure[\tiny][]{\label{t1}\includegraphics[width=7cm,height=5cm]{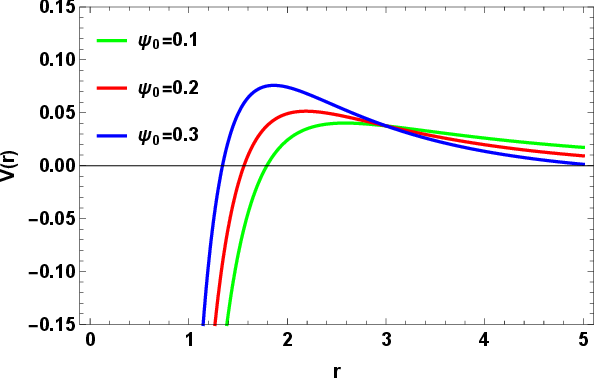}}\hfil
\subfigure[\tiny][]{\label{t2}\includegraphics[width=7cm,height=5cm]{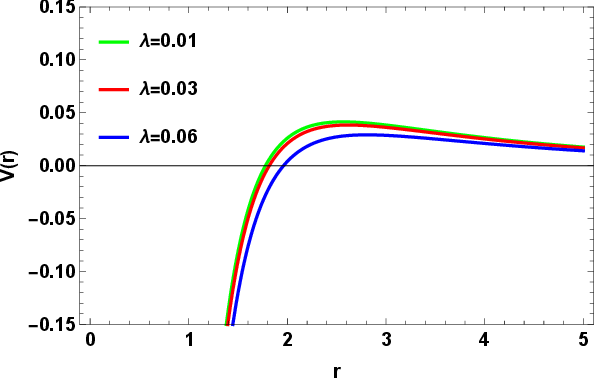}}
\caption{Plot showing the variation of $V(r)$ with respect to $r$
for different values of $\psi_{0}$ and $\lambda$ corresponding to
$\lambda=0.02$ (a) and $\psi_{0}=0.1$ (b) with $M=1$ and
$\Lambda=0.005$.}\label{evp3}
\end{figure}

A thorough grasp of unstable orbits is essential to understand the complex dynamics of particles
and the behavior of photons near the BH geometry. In the present analysis, these orbits play a pivotal role in
unravelling the intricacies of space-time influenced by the effects of the involved model parameters. To
obtain a comprehensive understanding of the impact on the photon sphere around the BH, we will utilize
the Lagrangian method to evaluate null geodesics. This method provides an easier path for readers
to comprehend the calculations as compared to those using the geodesic equation, as previously discussed.
We still use Eqs. (\ref{8})-(\ref{10}) and (\ref{14}) for the desired results. \footnote{Note that here and
everywhere below we use mathematical expressions without
dimensionless variables.} Now, we are interested in studying the most unstable photon orbits, and for that,
we must have to analyze the physical behavior of the effective potential $V(r)$ as shown in Fig. \textbf{\ref{evp3}}.
From Fig. \textbf{\ref{evp3}} (a), it is observed that the effective potential increases with the increasing values of
$\psi_{0}$ such as it is maximum when $\psi_{0}=0.4$. On the other hand, in Fig. \textbf{\ref{evp3}} (b), the effective
potential is decreasing with the increasing $\lambda$ values. These, in turn, affect the circular
orbits of the BH shadow.
The angular radius of the BH shadow that we consider in the present work is given by \cite{shd1,shd2}
\begin{equation}\label{17}
\sin^{2}\beta_{\textrm{sh}}=\frac{\mathcal{N}(r_{p2})^{2}}{\mathcal{N}(r_{obs})^{2}},
\end{equation}
with $\mathcal{N}(r)^{2}=r^{2}/A(r)$. The term $\beta_{\textrm{sh}}$
indicate the angular radius of the BH shadow, and $r_{obs}$ is
position of the distant observer. In addition, the quantity $r_{p2}$
is the radius of the photon orbit, as it was mentioned previously.
In this regard, Eq. (\ref{17}) can be further modify as
\begin{equation}\label{18}
\sin^{2}\beta_{\textrm{sh}}=\frac{r^{2}_{p2}}{A(r_{p2})}\frac{A(r_{obs})}{r^{2}_{obs}}.
\end{equation}
At a large distance, the radius of the BH shadow for an observer can be
determined as \cite{shd1}
\begin{eqnarray}\label{19}
R_{\textrm{sh}}=r_{obs}\sin\beta_{\textrm{sh}}=\frac{r_{p2}}{\sqrt{A(r_{p2})}}\sqrt{A(r_{obs})}.
\end{eqnarray}
To observe the BH shadow, we consider an observer located far away from the BH so that the neighborhood
of him can be viewed as asymptotically flat. After choosing a Cartesian coordinate system centered at the
BH, a closed curve will appear on the celestial coordinates, namely ($X,~Y$), resulting from the
projection of the unstable photon orbits. Hence, one can observe the radius and the shape of the BH
shadow via celestial coordinates \cite{shd3}, with the help of Eq. (\ref{19}).
\begin{figure}[H]\centering
\subfigure[\tiny][]{\label{t1}\includegraphics[width=5.2cm,height=5.2cm]{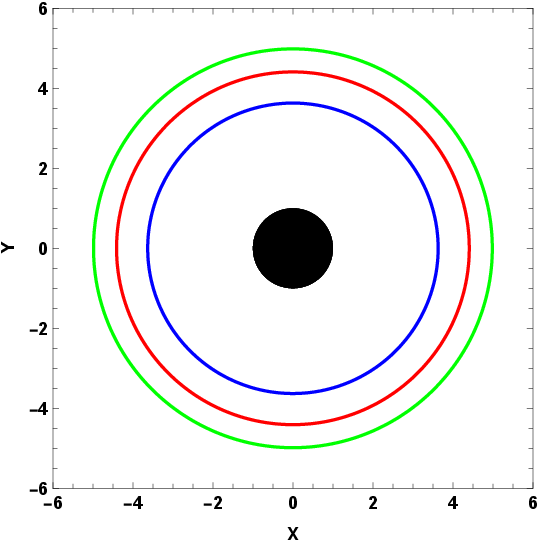}}\hfil
\subfigure[\tiny][]{\label{t2}\includegraphics[width=7cm,height=5cm]{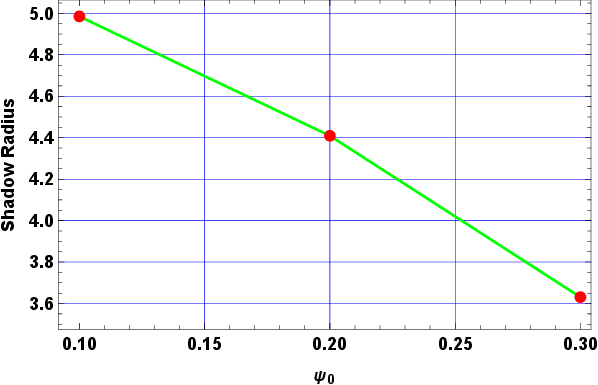}}
\caption{Figure (a) showing the shadow of BH for different values of
parameter $\psi_{0}=0.1$,~$0.2$,~$0.3$, for outer to inner circles
respectively, with $\lambda=0.02$,~$M=1$ and $\Lambda=0.005$. The
solid black disk shows the BH. Figure (b) illustrates the dependence
of shadow radius on the parameter $\psi_{0}$.}\label{evp4}
\end{figure}
\begin{figure}[H]\centering
\subfigure[\tiny][]{\label{t1}\includegraphics[width=5.2cm,height=5.2cm]{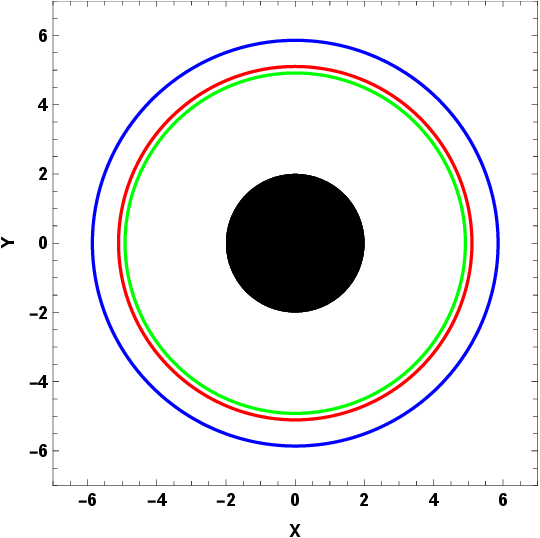}}\hfil
\subfigure[\tiny][]{\label{t2}\includegraphics[width=7cm,height=5cm]{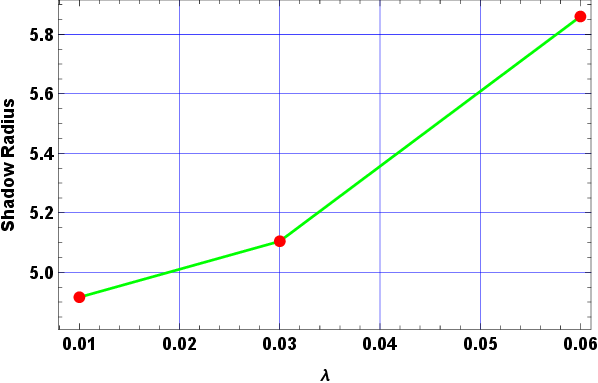}}
\caption{Figure (a) showing the shadow of BH for different values of
parameter $\lambda=0.01$,~$0.03$,~$0.06$, for inner to outer circles
respectively, with $\psi_{0}=0.1$,~$M=1$ and $\Lambda=0.005$. The
solid black disk shows the BH. Figure (b) illustrates the dependence
of shadow radius on the parameter $\lambda$.}\label{evp5}
\end{figure}

From Fig. \textbf{\ref{evp4}} (a), we observe that when the $f(R)$ parameter $\psi_{0}$ increases, the corresponding circles of the BH shadow move towards the central dark region. This effect can also
be observed from the corresponding BH shadow radius profile, where the blue line decreases with the
increase of the $\psi_{0}$, see Fig. \textbf{\ref{evp4}} (b). Moreover, the discrete red points
represent the corresponding specific values of $\psi_{0}$ used in plot \textbf{\ref{evp4}} (a), lie
on the green line. Besides, we analyze the BH shadow for different values of the monopole parameter
$\lambda$. Here, we observe that the effect of the parameter $\lambda$ on the radius of the BH
shadow is just opposite to that of the parameter $\psi_{0}$. As the increasing values of the parameter
$\lambda$ leads to an increase in the shadow radius, as shown in Fig. \textbf{\ref{evp5}} (a). An important different aspect is noted here that initially for smaller values of parameter $\lambda$, the radius of the BH shadow is slightly increased first and then suddenly increases with the aid of $\lambda$. The corresponding
profile of BH shadow radius also depicts the significance of the parameter $\lambda$ on the shadow of BH,
see Fig. \textbf{\ref{evp5}} (b). The individual red points lie on the green line interpret the corresponding
specific values of $\lambda$ used in plot \textbf{\ref{evp5}} (a). Hence, all these results indicate that
both parameters are sensitive to measuring the size of the BH shadow.

Now, we are going to discuss the dynamics of an optically sparse accretion flow that emits radiation around
the BH. We will employ a numerical method, the so-called backwards ray tracing procedure, to reveal the shadow image
generated by this radiation flow matter. In order to determine the intensity of the distribution matter within
the certain emitting region, we must consider the specific assumptions about the radiative and emission
mechanisms in play. In this perspective, one can observed the specific intensity $I({\upsilon_{o}})$
(usually measured in $\textrm{erg}{\textrm{s}}^{-1}{\textrm{cm}}^{-2}{\textrm{str}}^{-1}{\textrm{Hz}}^{-1}$)
at the point ($X, Y$) of the observer's image is given by \cite{shd4,shd5}
\begin{equation}\label{20}
I(\upsilon_{\textrm{o}})=\int_{\gamma} g^3
j(\upsilon_{e})dl_{\textrm{prop}},
\end{equation}
in which the symbol $\upsilon_{o}$,~$\upsilon_{e}$ is the observed and emitted photon frequency, respectively.
Further, the symbol $j(\upsilon_{e})$ is the emissivity per unit volume in the static frame, $\gamma$ is the
trajectory of emitted photons, $dl_{\textrm{prop}}$ is the infinitesimal proper length. Here, we consider the
case that the gas undergoes radial free fall. So, the red-shift factor $g$ can be defined as \cite{shd4}
\begin{equation}\label{21}
g=\frac {k_\mu u_{o}^\mu}{k_\nu u_{e}^\nu},
\end{equation}
where $u_{o}^\mu$,~$u_{e}^\nu$ is the four-velocity of the static observer and in-falling accretion matters,
respectively. For simplicity, we assume that the distant observer is stationary with $u_{o}^\mu=(1, 0, 0, 0)$.
While the four-velocity of the in-falling accretion matter is given by
\begin{eqnarray}\label{22}
u_e^t=\frac{1}{A(r)}; \quad \quad u_e^r=-\sqrt{1-A(r)}; \quad
u_e^\theta=u_e^\vartheta=0.
\end{eqnarray}
The term $k_\mu$ of Eq. (\ref{21}) is the four-momentum of the photon, radiated from accretion matter,
which can be calculated via $k_\mu=\partial\mathcal{L}/\partial\dot{x}^{\mu}$. Importantly, both
$u_{o}^\mu$ and $u_{e}^\nu$ are the only components of $t$ or $r$, thus we only calculate the relationship
between $k_{r}$ and $k_{t}$, as given below
\begin{eqnarray}\label{23}
k_{t}=\frac{1}{b}, \quad
k_{r}=\pm\frac{1}{A(r)b}\sqrt{1-\frac{b^{2}A(r)}{r^{2}}},
\end{eqnarray}
The sign $+(-)$ indicates the direction in which the photon is either approaching or moving away from
the massive object, respectively. Based on Eqs. (\ref{21})-(\ref{23}), the red-shift factor can be defined as
\begin{equation}\label{24}
g=\frac{1}{\bigg(u_{e}^{t}\pm\frac{k_{r}}{k_{t}}u_{e}^{r}\bigg)}.
\end{equation}
In addition, regarding the specific emissivity, we employ a straightforward model in which the emission
is monochromatic, having an emitter's static frame frequency denoted as $\nu_{\ast}$ and a $\frac{1}{r^{2}}$
radial profile
\begin{equation}\label{25}
j(\upsilon_e)\propto\frac{\delta(\upsilon_e-\nu_{\ast})}{r^2},
\end{equation}
in which $\delta$ is the Dirac delta function, along with the infinitesimal proper length, which is
\begin{equation}\label{26}
dl_{\textrm{prop}}=\frac{k_{t}}{g|k_{r}|}dr.
\end{equation}
Integrating the intensity across all observable frequencies, we arrive at the observed flux for the
in-falling spherical flow matter as follows
\begin{equation}\label{27}
I_{\textrm{obs}}(\upsilon_{\textrm{o}})\propto\int_{\gamma}
\frac{g^3 k_t}{ r^2|k_{r}|}dr.
\end{equation}
\begin{figure}[H]\centering
\includegraphics[width=5.3cm,height=4.5cm]{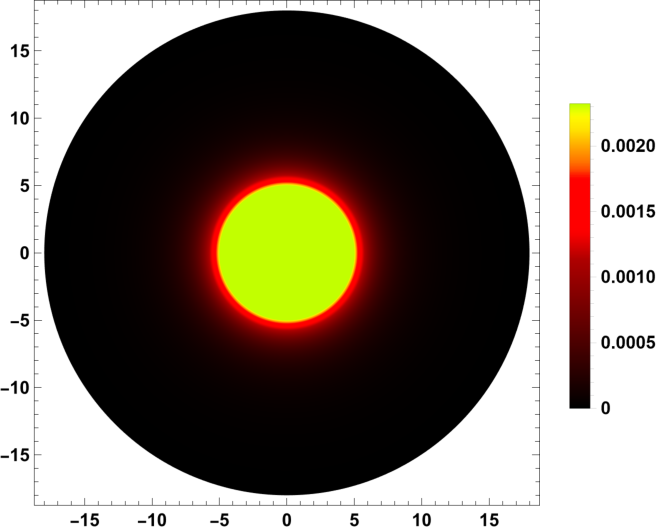}
\includegraphics[width=5.3cm,height=4.5cm]{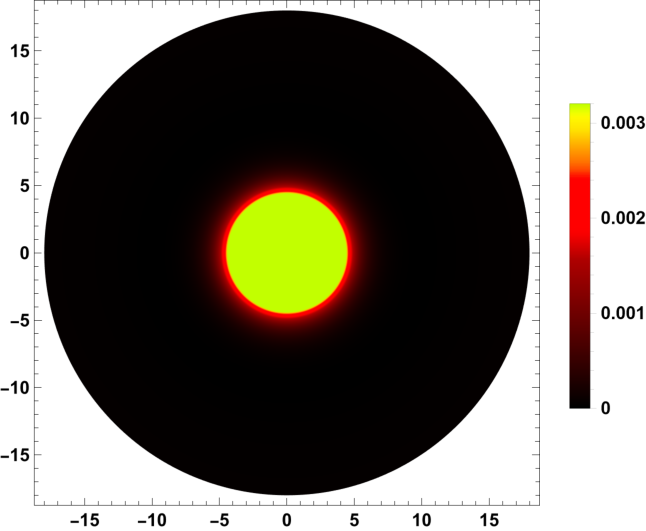}
\includegraphics[width=5.3cm,height=4.5cm]{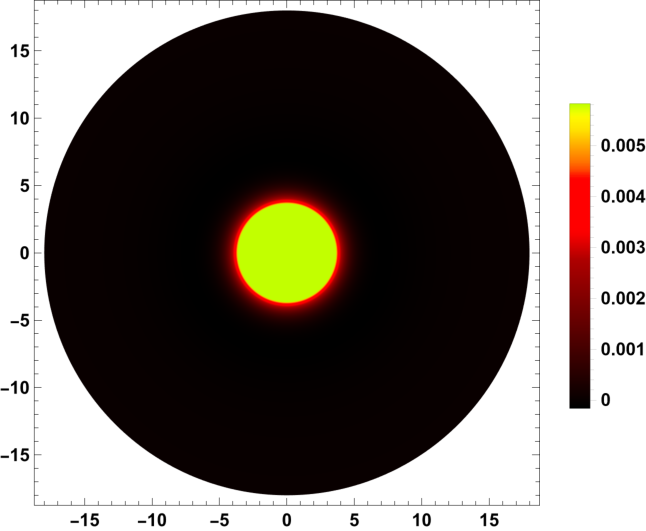}
\caption{Optical signatures of the observed specific intensity using
the radially infalling accretion model for different values of
parameter $\psi_{0}$. We plot these two-dimensional intensity maps
in celestial coordinates and from left to right panels, the values of
$\psi_{0}$ corresponds to $\psi_{0}=0.1$,~$0.2$,~$0.3$,
respectively, with $\lambda=0.02$,~$M=1$ and $\Lambda=0.005$, as an
three examples.}\label{fal1}
\end{figure}
\begin{figure}[H]\centering
\includegraphics[width=5.3cm,height=4.5cm]{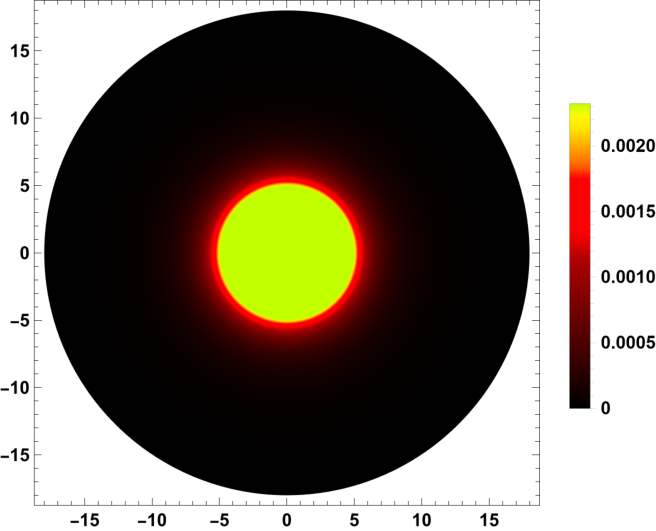}
\includegraphics[width=5.3cm,height=4.5cm]{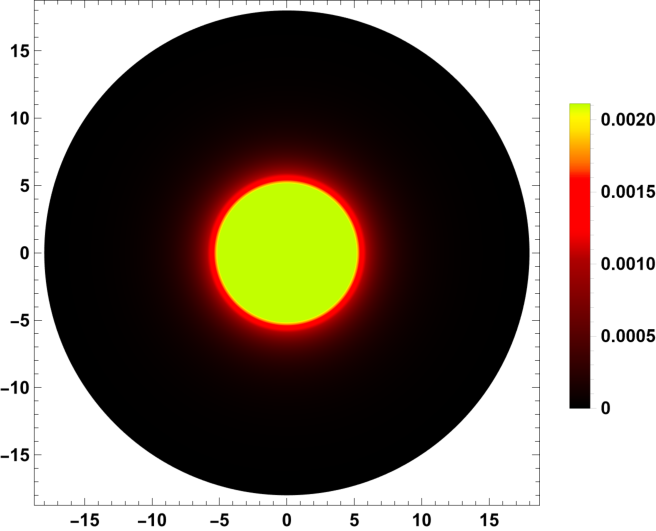}
\includegraphics[width=5.3cm,height=4.5cm]{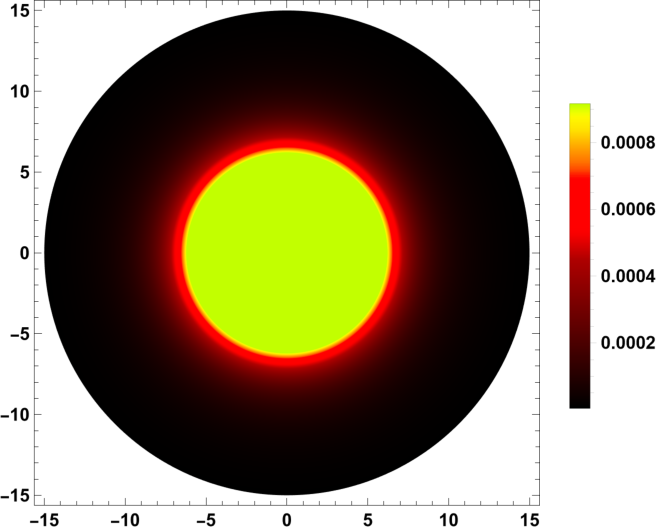}
\caption{Optical signatures of the observed specific intensity using
the radially infalling accretion model for different values of
parameter $\lambda$. We plot these two-dimensional intensity maps
in celestial coordinates and from left to right panels, the values of
$\lambda$ corresponds to $\lambda=0.01$,~$0.03$,~$0.06$,
respectively, with $\psi_{0}=0.1$,~$M=1$ and $\Lambda=0.005$, as an
three examples.}\label{fal2}
\end{figure}

In Figs. {\bf\ref{fal1}} and {\bf\ref{fal2}}, the optical appearance
of infalling accretion matter is presented for different values of
$\psi_{0}$ and $\lambda$, respectively. Particularly, in each panel,
there is a bright circular ring in the middle of the screen, which
corresponds to the position of the photon sphere at $b=b_{c}$. Clearly,
it can be observed that increasing the value of $\psi_{0}$ leads to
a decrease in the shadow radius, while it increases with the
increasing of $\lambda$. These trends are also evident in Figs.
\textbf{\ref{evp4}} and \textbf{\ref{evp5}}, where the circular
orbits indicate the location of the photon sphere.

\section{Consequences of Thermal Fluctuations of $f(R)$ AdS Black Hole}

This segment is elaborated to explore the graphical development of various important
measures like entropy, corrected pressure, Van der Waals file, specific heat, internal energy,
enthalpy under the thermal fluctuations of $f(R)$ AdS BH.

\subsection{Corrected Entropy}

In the context of high-energy physics, the influence of quantum fluctuations is essential and can be
regarded as an adjustment to the infrared limit which further plays an important role in maintaining
the stability of BH as well as high energy limit of thermodynamical measures \cite{SN25}. It is
well-known that the BH entropy can computed via relationship: $S_o=\frac{A}{4}$ with $A$ referring to
BH event horizon area. In the current setup, the corrected entropy is given by \cite{SN26,SN13}
\begin{eqnarray}\nonumber
S=\pi r^2-\frac{\sigma}{2}\ln\big[\pi
r^2\left(\frac{2-16\pi\lambda^2+2r^2\Lambda-4r\psi_0-2r^3\Lambda\psi_0
+3r^2\psi_0^2}{(8\pi r-12\pi r^2\psi_0)}\right)^2\big].
\end{eqnarray}
\begin{figure}[H]\centering
\includegraphics[width=5.3cm,height=5.0cm]{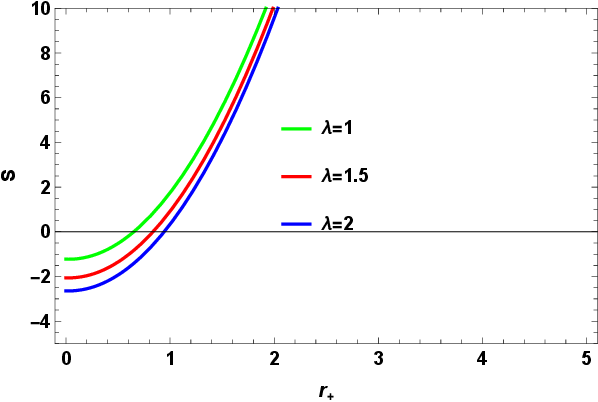}
\includegraphics[width=5.3cm,height=5.0cm]{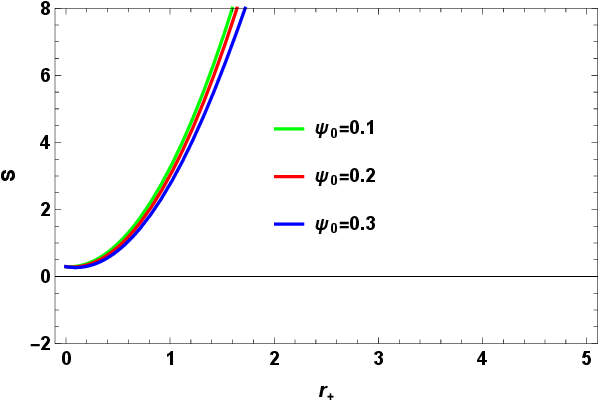}
\includegraphics[width=5.3cm,height=5.0cm]{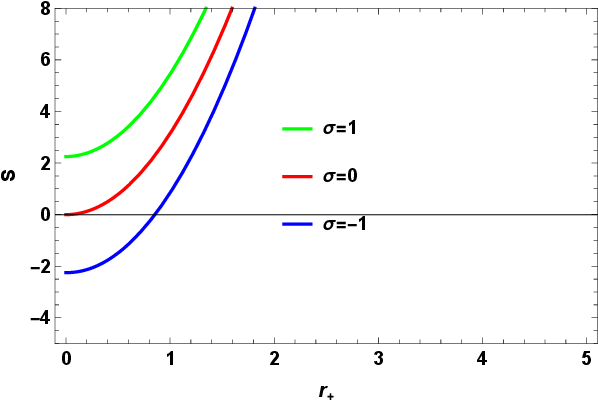}
\caption{Effects of $\lambda$ with fixed $\psi_{0}=0.1$ and
$\sigma=1$ (left panel), $\psi_{0}$ with fixed $\lambda=0.5$ and
$\sigma=1$ (middle panel) and $\sigma$ with fixed $\lambda=0.1$ and
$\psi_{0}=0.05$ (right panel) on corrected entropy. In all cases, we
fixed $\Lambda=0.005$.}\label{thm1}
\end{figure}
Figure \textbf{\ref{thm1}} demonstrates the graphical response of
corrected entropy of $f(R)$ AdS BH against radius of event horizon
when different choices of model parameters $\lambda,~ \psi_0$ and
$\sigma$ are taken. It is obvious from these graphs that the BH
radius and entropy exhibit directly proportional relation as one
quantity increases with the increase in other quantity. From the
left panel of Fig. \textbf{\ref{thm1}}, it is seen that for small
domain of horizon radius, the entropy exhibits negative trend and
hence refers to instability as well as phase transition. It can also
be checked that for decreasing $\lambda$ values, the entropy is
increasing with $r_{+}$. In the middle plot, it can be noticed that
the entropy exhibits positive trend for all values of horizon radius
and the entropy increases with decreasing $\psi_0$ values, which
shows the stable behavior. The right panel of Fig.
\textbf{\ref{thm1}} shows that negative $\sigma$ along with small
horizon radius correspond to negative trend of entropy and hence
indicates instability and phase transition there. For $\sigma=0$,
the entropy exhibits exponential like behavior while positive
$\sigma$ values indicates linearly increasing behavior supporting
the BH stability.

\subsection{Helmholtz Free Energy}

\begin{figure}[H]\centering
\includegraphics[width=5.3cm,height=5.0cm]{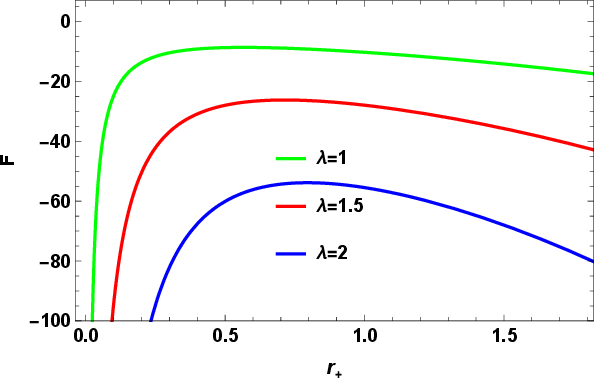}
\includegraphics[width=5.3cm,height=5.0cm]{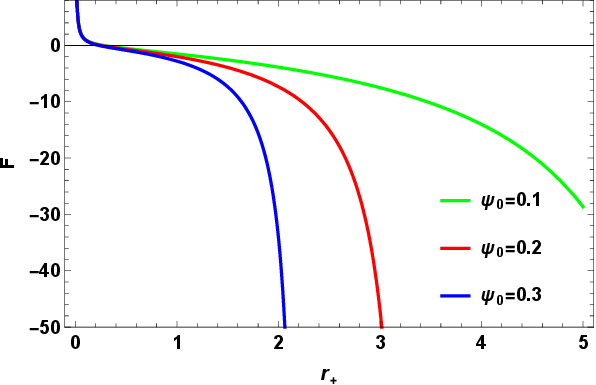}
\includegraphics[width=5.3cm,height=5.0cm]{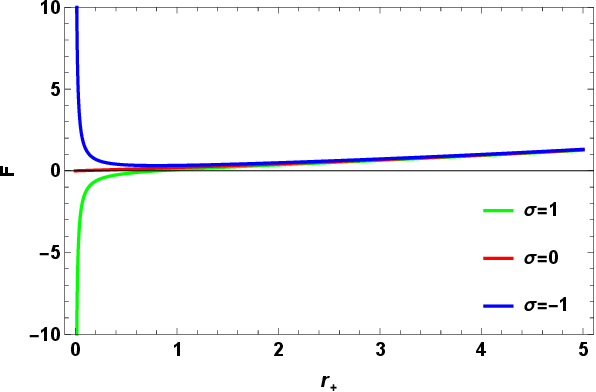}
\caption{Effects of $\lambda$ with fixed $\psi_{0}=0.1$ and
$\sigma=1$ (left panel), $\psi_{0}$ with fixed $\lambda=0.5$ and
$\sigma=1$ (middle panel) and $\sigma$ with fixed $\lambda=0.1$ and
$\psi_{0}=0.05$ (right panel) on Helmholtz free energy. In all
cases, we fixed $\Lambda=0.005$.}\label{thm2}
\end{figure}

In this segment, we shall talk about the graphical response of Helmholtz free energy (HFE) for
$f(R)$ AdS BH. It is argued that HFE is a fundamental quantity in BH thermodynamics as it governs
the stability and equilibrium properties of BH. More specifically, HFE is linked to the first thermodynamic
law of BH, which describes how variations in BH mass relate to changes in entropy, volume, and other parameters,
such as the cosmological constant. Mathematically, it can be written as $F=M-TS$ which for $f(R)$ AdS BH
takes the form below:
\begin{eqnarray}\nonumber
F&=&\frac{r(3-24\pi\lambda^2+r^2\Lambda-3r\psi_0)}{6-9r\psi_0}-
\frac{-2+16\pi\lambda^2+r(4\psi_0-3r\psi_0^2+2r\Lambda(-1+r\psi_0))}{8\pi r(-2+3r\psi_0)}\\\nonumber&\times&
\left(2\pi r^2+\sigma \ln[16\pi]-\sigma\ln\big[\frac{(-2+16\pi\lambda^2+r(4\psi_0-3r\psi_0^2+2r\Lambda(-1+r\psi_0)))^2}{(2-3r\psi_0)^2}\big]\right).
\end{eqnarray}
Figure \textbf{\ref{thm2}} illustrates the behavior of the HFE for
the $f(R)$ AdS BH as a function of the event horizon $r_+$ under
different variations of the model parameters. From the left panel,
it is evident that the HFE remains negative for all considered
values of $\lambda$, which indicates thermodynamic stability and the
possible occurrence of a phase transition. The middle panel of Fig.
\textbf{\ref{thm2}} shows that increasing $\psi_{0}$ leads to
positive values of the HFE only within a small range of the horizon
radius. This behavior signifies the presence of a phase transition
within that limited domain. However, with the aid of $r_+$, the HFE
shows negative values, which is corresponds to the thermodynamic
stability. The impact of $\sigma$ on the HFE are depicted in the
right panel of Fig. \textbf{\ref{thm2}}. Here, it is observed that,
when $\sigma=-1$, the HFE attains positive values and does not
exhibit any phase transition. In contrast, for $\sigma=1$, the HFE
initially takes negative values, indicating a stable phase, and then
undergoes a phase transition around $r_+\approx0.9$ after which it
becomes positive. At $\sigma=0$, the HFE remains $F=0$ for small
values of $r_+$. As $r_+$ increases, the HFE becomes positive,
indicating that the BH configuration is thermodynamically unstable.
Overall, it is clear that the effects of $\lambda$ and $\psi_{0}$ on
the HFE are qualitatively similar, whereas the parameter $\sigma$
introduces a distinct thermodynamic behavior, particularly in
relation to phase transitions at a fixed horizon radius.

\subsection{Internal Energy}

Herein, we shall provide the graphical illustration of internal energy for $f(R)$ AdS BH.
Generally, the internal energy for a BH configuration can be evaluated through the relation: $E=F+TS$.
In the present setup, the respective expression can be written as
\begin{eqnarray}\nonumber
E=\frac{r(3-24\pi\lambda^2+r^2\Lambda-3r\psi_0)}{6-9r\psi_0}.
\end{eqnarray}
\begin{figure}[H]\centering
\includegraphics[width=6.3cm,height=5.0cm]{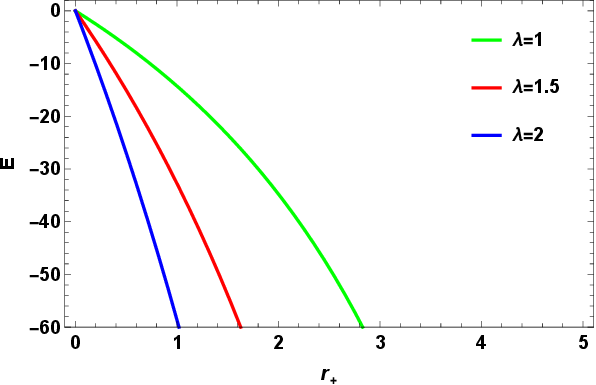}
\includegraphics[width=6.3cm,height=5.0cm]{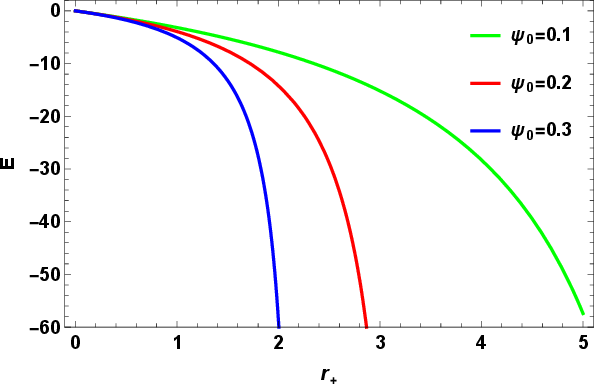}
\caption{Effects of $\lambda$ with fixed $\psi_{0}=0.1$ (left panel)
and $\psi_{0}$ with fixed $\lambda=0.5$ (right panel) on internal
energy. In all cases, we fixed $\Lambda=0.005$.}\label{thm3}
\end{figure}
From the plots of Fig. \textbf{\ref{thm3}}, it is easy to check that
for both variations of parameters $\lambda$ and $\psi_0$, the
internal energy turns out as a negative quantity for all domain of
event horizon and thus it can be concluded that in present setup,
the internal energy shows inversely decreasing behavior with radius
and hence refers to instability. Also, the smaller values of both
$\lambda$ and $\psi_0$ parameters, the internal energy $E$
increasing behavior (through negative values) with respect to
$r_{+}$. Near $r_{+}=0$, the lower energy suggests strong attractive
or stable interactions, but as $r_{+}$ increases, the system shows
instability. This consistent shape also reflects an instability
potential well, where the system naturally tends to settle at the
energy minimum.

\subsection{Corrected Pressure}

In this part, we shall graphically explore the behavior of corrected pressure for $f(R)$ AdS BH.
Mathematically, for a given BH configuration, the corrected pressure can be found by using the relation:
$P=-\frac{dF}{dV}$. In the current scenario, the value of corrected pressure is given by
\begin{eqnarray}\nonumber
P&=&-\frac{1}{16(2-3r\psi_0)^2}\bigg[r(-2\pi r(-2+16\pi\lambda^2+r^2(-9\psi_0^2(-1+r\psi_0)
+\Lambda(2+r\psi_0(-10+9r\psi_0))))\\\nonumber&+&(-2(\psi_0+24\pi\lambda^2\psi_0)+r(8\Lambda+3\psi_0^2(4-3r\psi_0)
+6r\Lambda\psi_0(-3+2r\psi_0)))\sigma)+(2+16\pi\lambda^2(-1+3r\psi_0)\\\nonumber&+&r(-6\psi_0+3r\psi_0^2+r\Lambda
(-2+r\psi_0(4-3r\psi_0))))\sigma\ln[16\pi]+(-2+16\pi\lambda^2(1-3r\psi_0)+r(6\psi_0-3r\psi_0^2\\\nonumber&+&r\Lambda
(2+r\psi_0(-4+3r\psi_0))))\sigma\ln[\frac{(-2+16\pi\lambda^2+r(4\psi_0-3r\psi_0^2+2r\Lambda)(-1+r\psi_0))^2}{(2-3r\psi_0)^2}]\bigg].
\end{eqnarray}
\begin{figure}[H]\centering
\includegraphics[width=5.3cm,height=5.0cm]{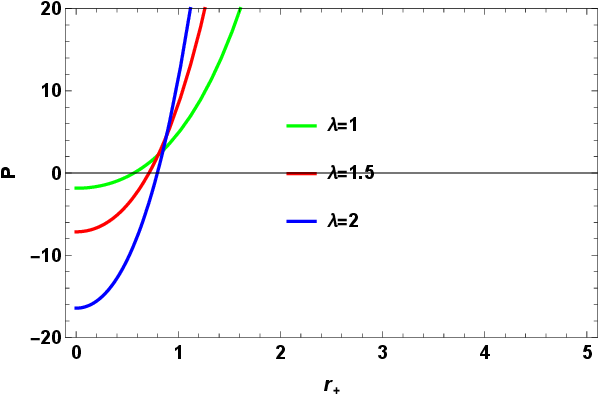}
\includegraphics[width=5.3cm,height=5.0cm]{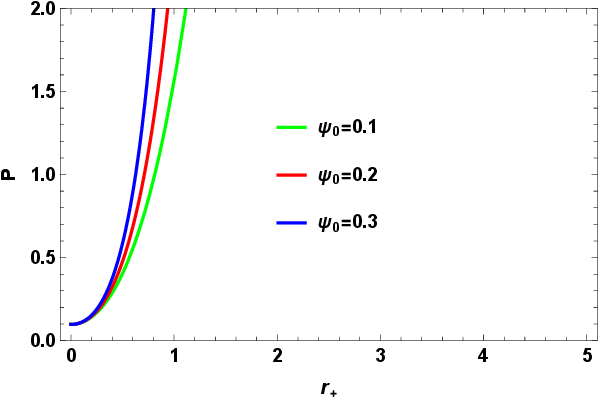}
\includegraphics[width=5.3cm,height=5.0cm]{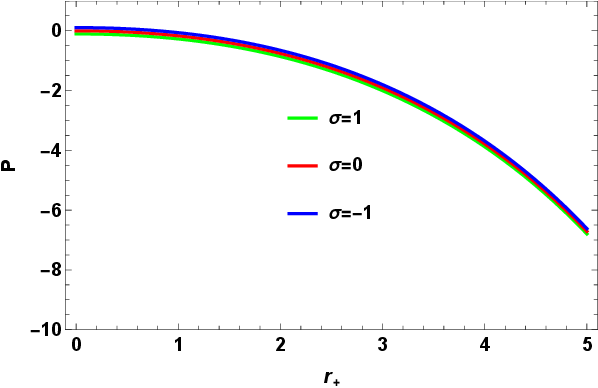}
\caption{Effects of $\lambda$ with fixed $\psi_{0}=0.1$ and
$\sigma=1$ (left panel), $\psi_{0}$ with fixed $\lambda=0.5$ and
$\sigma=1$ (middle panel) and $\sigma$ with fixed $\lambda=0.1$ and
$\psi_{0}=0.05$ (right panel) on corrected pressure. In all cases,
we fixed $\Lambda=0.005$.}\label{thm4}
\end{figure}
Figure \textbf{\ref{thm4}} provides the graphical examination of
corrected pressure $P$ for $f(R)$ AdS BH versus $r_+$. From the left
panel, it can be observed that after $r_+\approx1$, the corrected
pressure exhibits positive behavior regardless of whether it is small or large
$\lambda$ values. However, for $r\lesssim1$, it is seen that small
$\lambda$ values refer to increasing pressure values, while after
$r_+\approx1$, the large $\lambda$ variations produce rapidly
increasing corrected pressure. In the middle plot, it is obvious
that the corrected pressure exhibits positive and directly
proportional relationship (both quantities increase together) in the
whole domain of horizon radius. In the right plot, it can be checked
that for all values of $\sigma$ with respect to $r_+$, the corrected
pressure turns out to be negative. It can also be seen that negative
$\sigma$ values may produce a positive trend of corrected pressure for
small horizon radius. The comparison of all plots shows that the
corrected pressure remains positive under variations of $\lambda$
and $\psi_{0}$, while it becomes negative with $\sigma$. This
suggests that the contributions from $\lambda$ and $\psi_{0}$ are
more prominent than those from $\sigma$. At $P=0$, the system
exhibits critical behavior at certain values of the horizon radius
$r_+$, indicating the onset of thermodynamic stability.

\subsection{Corrected Enthalpy}

Herein, we shall evaluate the corrected enthalpy for the considered
BH and examine the corresponding expression graphically. Enthalpy is
another notable thermodynamic quantity, as it provides fruitful
insight into stability changes during thermodynamic processes and
also helps in distinguishing the equilibrium conditions of a
configuration. Its role in BH thermodynamics has gained considerable
significance following its incorporation into the first law of
thermodynamics. For a BH configuration, the mass represents the
enthalpy, encompassing not only the internal energy but also the
energy required to assemble the system along with its surrounding
environment. In this sense, the mass of an AdS BH functions as
enthalpy, analogous to its role in classical thermodynamics.
Generally, enthalpy can be computed through the relation: $H=E+PV$.
For $f(R)$ AdS BH, the resulting expression is given by
\begin{eqnarray}\nonumber
H&=&\frac{r(3-24\pi\lambda^2+r^2\Lambda-3r\psi_0)}{6-9r\psi_0}-\frac{\pi r^3}{12(2-3r\psi_0)^2}
\left(r(-2\pi r(-2+16\pi\lambda^2+r^2(-9\psi_0^2(-1+r\psi_0)+\Lambda\right.\\\nonumber&\times&\left.(2+r\psi_0(-10+9r\psi_0))))
+(-2(\psi_0+24\pi\lambda^2\psi_0)+r(8\Lambda+3\psi_0^2(4-3r\psi_0)+6r\Lambda\psi_0(-3+2r\psi_0)))\sigma)+(2\right.\\\nonumber&+&\left.
16\pi\lambda^2(-1+3r\psi_0)+r(-6\psi_0+3r\psi_0^2+r\Lambda(-2+r\psi_0(4-3r\psi_0))))\sigma
\ln[16\pi]+(-2+16\pi\lambda^2(1-3r\psi_0)\right.\\\nonumber&+&\left.r(6\psi_0-3r\psi_0^2+r\Lambda(2+r\psi_0(-4+3r\psi_0))))\sigma
\ln[\frac{(-2+16\pi\lambda^2+r(4\psi_0-3r\psi_0^2+2r\Lambda(-1+r\psi_0)))^2}{(2-3r\psi_0)^2}]\right).
\end{eqnarray}
\begin{figure}[H]\centering
\includegraphics[width=5.3cm,height=5.0cm]{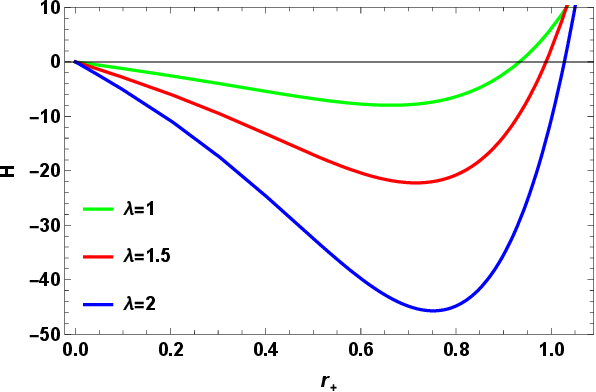}
\includegraphics[width=5.3cm,height=5.0cm]{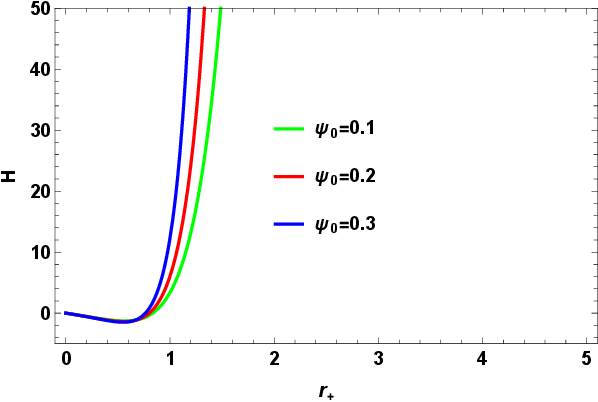}
\includegraphics[width=5.3cm,height=5.0cm]{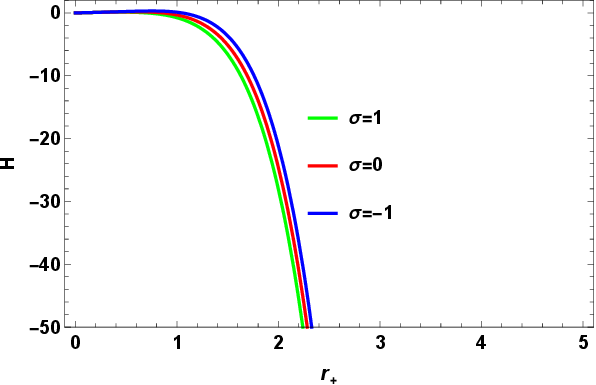}
\caption{Effects of $\lambda$ with fixed $\psi_{0}=0.1$ and
$\sigma=1$ (left panel), $\psi_{0}$ with fixed $\lambda=0.5$ and
$\sigma=1$ (middle panel) and $\sigma$ with fixed $\lambda=0.1$ and
$\psi_{0}=0.05$ (right panel) on corrected enthalpy. In all cases,
we fixed $\Lambda=0.005$.}\label{thm5}
\end{figure}

The graphical behavior of the corrected enthalpy $H$ is presented in
Fig.~\textbf{\ref{thm5}}. From the left panel, it is observed that
the enthalpy exhibits a negative trend with variations in $\lambda$
for $r_+ \lesssim 1$, indicating unstable configurations. Moreover,
for smaller values of $\lambda$, the enthalpy shows an increasing
behavior with respect to the horizon radius. From the middle panel,
it is evident that the enthalpy increases with the increasing
values of $\psi_0$. This quantity becomes negative in the region $0
< r_+ < 1$ for different values of $\psi_0$, whereas for $r_+ > 1$,
it attains positive values. In the right panel, for different
variations of $\sigma$, the enthalpy remains negative over the
entire domain of the horizon radius (see right panel). Additionally,
all these results indicate that the impact of $\lambda$ is more
prominent on the physical behavior of enthalpy, as compared to
$\psi_0$ and $\sigma$, and system is stable and non-chaotic with
respect to $\sigma$.

\subsection{Gibbs Free Energy}

\begin{figure}[H]\centering
\includegraphics[width=5.3cm,height=5.0cm]{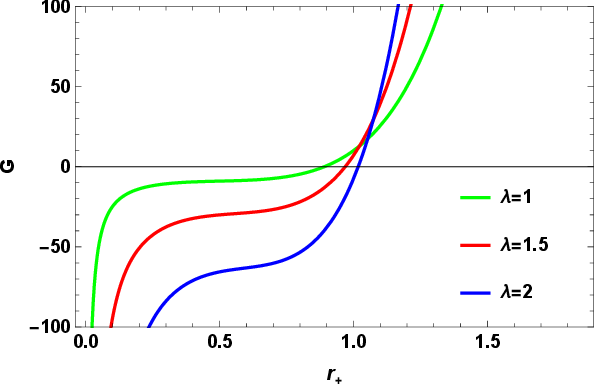}
\includegraphics[width=5.3cm,height=5.0cm]{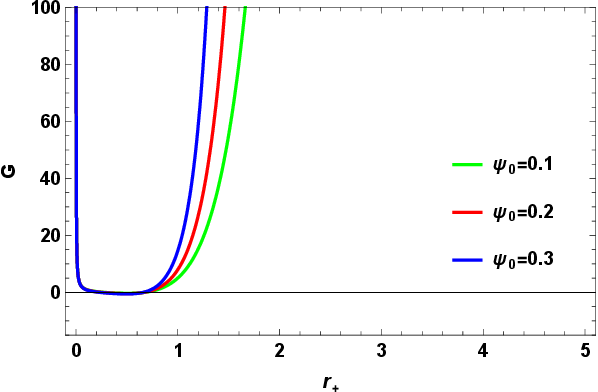}
\includegraphics[width=5.3cm,height=5.0cm]{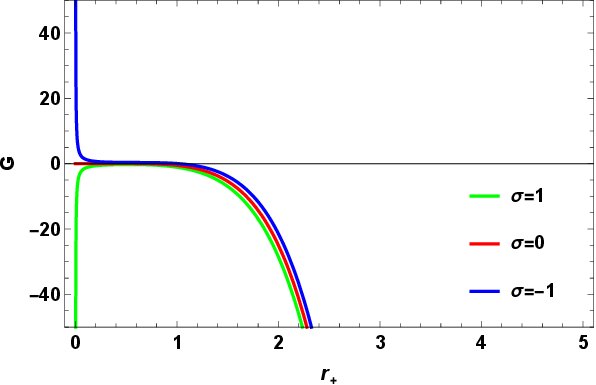}
\caption{Effects of $\lambda$ with fixed $\psi_{0}=0.1$ and
$\sigma=1$ (left panel), $\psi_{0}$ with fixed $\lambda=0.5$ and
$\sigma=1$ (middle panel) and $\sigma$ with fixed $\lambda=0.1$ and
$\psi_{0}=0.05$ (right panel) on Gibbs free energy. In all cases, we
fixed $\Lambda=0.005$.}\label{thm6}
\end{figure}

Herein, we shall investigate the graphical development of Gibbs free
energy (GFE) for $f(R)$ AdS BH. To maintain a constant boundary for
$f(R)$ AdS BH at a constant temperature, both the pressure and
temperature must be held fixed. Under these conditions, the
appropriate thermodynamic potential is the GFE. Once the enthalpy,
HFE, pressure, and volume are specified; the GFE can be determined
by utilizing the relation: $G=F+PV$. In the present case, the respective
expression can be written as
\begin{eqnarray}\nonumber
G&=&\frac{r(3-24\pi\lambda^2+r^2\Lambda-3r\psi_0)}{6-9r\psi_0}-\frac{-2+16\pi\lambda^2
+r(4\psi_0-3r\psi_0^2+2r\Lambda(-1+r\psi_0))}{8\pi r(-2+3r\psi_0)}\left(2\pi r^2+\sigma
\ln[16\pi]\right.\\\nonumber&-&\left.\sigma\ln[\frac{-2+16\pi\lambda^2+r(4\psi_0-3r\psi_0^2
+2r\Lambda(-1+r\psi_0))^2}{(2-3r\psi_0)^2}]\right)-\frac{\pi r^3}{12(2-3r\psi_0)^2}\left(r(-2\pi r(-2+16\pi\lambda^2\right.\\\nonumber&+&\left.r^2(-9\psi_0^2(-1+r\psi_0)+\Lambda(2+r\psi_0
(-10+9r\psi_0))))+(-2(\psi_0+24\pi\lambda^2\psi_0)+r(8\Lambda+3\psi_0^2(4-3r\psi_0)\right.\\\nonumber&+&\left. 6r\Lambda\psi_0(-3+2r\psi_0)))\sigma)+(2+16\pi\lambda^2(-1+3r\psi_0)+r(-6\psi_0+3r\psi_0^2+r\Lambda(-2+r\psi_0(4-3r\psi_0))))
\right.\\\nonumber&\times&\left.\sigma\ln[16\pi]
+(-2+16\pi\lambda^2(1-3r\psi_0)+r(6\psi_0-3r\psi_0^2+r\Lambda(2+r\psi_0(-4+3r\psi_0))))\sigma\right.\\\nonumber&\times&\left.
\ln[\frac{(-2+16\pi\lambda^2+r(4\psi_0-3r\psi_0^2+2r\Lambda(-1+r\psi_0)))^2}{(2-3r\psi_0)^2}]\right).
\end{eqnarray}
The graphical behavior of the GFE is illustrated in
Fig.~\textbf{\ref{thm6}}. From the left panel, it is observed that
for $r_+ < 1$, the GFE attains negative values, indicating that the
BH phase is thermodynamically favored in this region. However, for
$r_+ > 1$, the GFE becomes positive for all variations of $\lambda$,
suggesting that the corresponding configurations are
thermodynamically unstable. It is further noted that the GFE
increases with increasing values of $\lambda$, and there is a phase
transition between globally stable configurations to unstable
regime. The middle panel shows that the GFE remains positive for all
values of $\psi_0$ and exhibits a characteristic U-shaped profile
with a distinct minimum. This behavior signifies that the system is
thermodynamically unstable at both small and large horizon radii,
while the minimum point corresponds to a relatively more stable
equilibrium configuration where the free energy is minimized. From
the right panel, it is evident that for $\sigma = 0$ and $\sigma =
1$, the GFE remains negative throughout the entire range of the
horizon radius, indicating that the BH phase is globally stable and
thermodynamically preferred in these cases. In contrast, for $\sigma
=-1$, the GFE becomes positive in the small horizon radius regime,
implying that such configurations are thermodynamically unstable,
whereas stability may be achieved as the horizon radius increases
and the GFE decreases.

\subsection{Change in Corrected Pressure}

In this segment, we shall calculate the change in pressure for the
considered BH geometry through van der Waals' system. The van der
Waals system is one of the significantly applicable models for
conversion between liquid and gas phases. It is pointed out in
the literature that using the corrected ideal gas equation, one can
derive the equation of state for this model. Mathematically, the van
der Waals model is defined by the equation below \cite{stvg}
\begin{equation}\nonumber
T(P,v)K=\left(P_v+\frac{a}{v^2}\right)(V-B).
\end{equation}
Herein, the notations $a,~ B$ and $K$, respectively, refer to the
size, molecule attraction and Boltzmann constant. The analogy
between fluid and BH temperatures is clarified using the van der
Waals BH correspondence. This model can be re-written as
\begin{equation}\nonumber
P_v=\frac{T}{V-B}-\frac{a}{v^2}.
\end{equation}
In the current setup, it can be written as
\begin{eqnarray}\nonumber
P_v=-\frac{a}{v^2}+\frac{-6+48\pi\lambda^2+3r(4\psi_0-3r\psi_0^2+2r\Lambda(-1+r\psi_0))}{4\pi
r(-3B+4\pi r^3)(-2+3r\psi_0)}.
\end{eqnarray}
Consequently, the pressure change can be computed via the relation $\nabla P=P-P_v$ and is given by
\begin{eqnarray}\nonumber
\nabla
P&=&\frac{a}{v^2}-\frac{-6+48\pi\lambda^2+3r(4\psi_0-3r\psi_0^2+2r\Lambda(-1+r\psi_0))}{4\pi
r(-3B+4\pi r^3)(-2+3r\psi_0)}
-\frac{1}{16(2-3r\psi_0)^2}\left(r(-2\pi
r(-2+16\pi\lambda^2\right.\\\nonumber
&+&\left.r^2(-9\psi_0^2(-1+r\psi_0)+\Lambda(2+r\psi_0(-10+9r\psi_0))))
+(-2(\psi_0+24\pi\lambda^2\psi_0)+r(8\Lambda+3\psi_0^2(4-3r\psi_0)\right.\\\nonumber
&+&\left.6r\Lambda\psi_0(-3+2r\psi_0)))\sigma)+(2+16\pi\lambda^2(-1+3r\psi_0)
+r(-6\psi_0+3r\psi_0^2+r\Lambda(-2+r\psi_0(4-3r\psi_0))))\right.\\\nonumber
&\times&\left.\sigma\ln[16\pi]+(-2+16\pi\lambda^2(1-3r\psi_0)+r(6\psi_0-3r\psi_0^2+r\Lambda
(2+r\psi_0(-4+3r\psi_0))))\sigma\right.\\\nonumber
&\times&\left.\ln[\frac{(-2+16\pi\lambda^2+r(4\psi_0-3r\psi_0^2+2r\Lambda(-1+r\psi_0)))^2}{(2-3r\psi_0)^2}]\right).
\end{eqnarray}
\begin{figure}[H]\centering
\includegraphics[width=5.3cm,height=5.0cm]{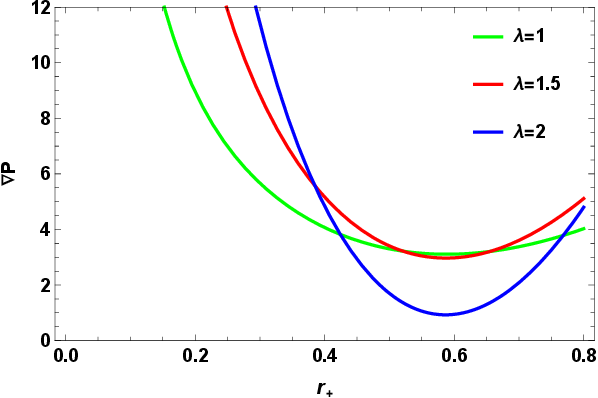}
\includegraphics[width=5.3cm,height=5.0cm]{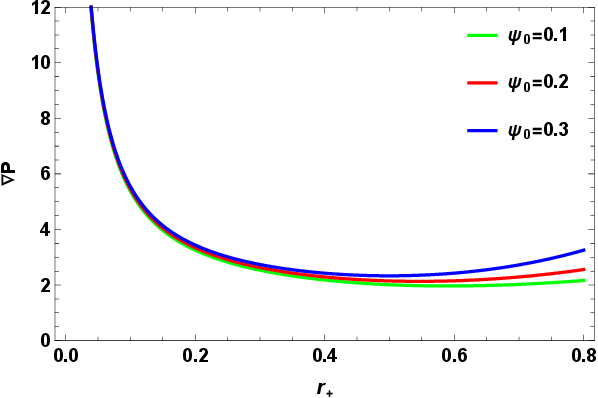}
\includegraphics[width=5.3cm,height=5.0cm]{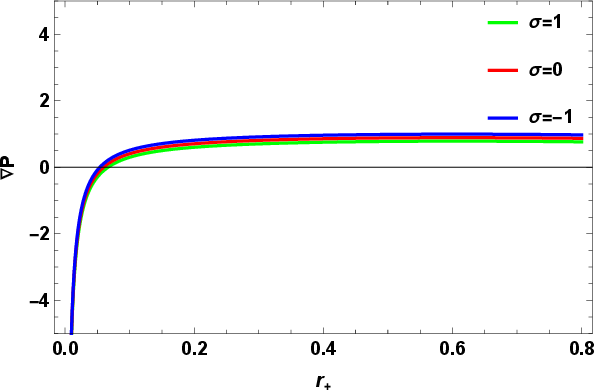}
\caption{Effects of $\lambda$ with fixed $\psi_{0}=0.1$ and
$\sigma=1$ (left panel), $\psi_{0}$ with fixed $\lambda=0.5$ and
$\sigma=1$ (middle panel) and $\sigma$ with fixed $\lambda=0.1$ and
$\psi_{0}=0.05$ (right panel) on change in corrected pressure. In
all cases, we fixed $\Lambda=0.005$,~$a=\eta=1$ and
$B=-1$.}\label{thm7}
\end{figure}
The expression for change in corrected pressure $\nabla P$ has been
analyzed in the Fig.~\textbf{\ref{thm7}}. From the left and middle
panels, it is noticed that the corrected pressure exhibits a positive
and decreasing trend for both $\lambda$ and $\psi_0$ variations with
increasing horizon radius. It is also obvious that corrected
pressure may exhibit increasing behavior after $r_+\approx0.8$ for
all considered variations of these parameters. From the right panel,
it is observed that $\nabla P$ initially takes negative values
around $r_+ \approx 0.1$, indicating a decreasing pressure profile
in the small horizon radius regime. As $r_+$ increases, a transition
point is encountered where $\nabla P$ changes sign and becomes
positive for different values of $\sigma$, suggesting a change in
the thermodynamic behavior of the system. Beyond this transition
region, $\nabla P$ tends to approach a nearly constant profile with
increasing $r_+$. Furthermore, all panels demonstrate that
variations in $\lambda$ have a more pronounced effect on $\nabla P$
compared to $\psi_0$ and $\sigma$, indicating that $\lambda$ plays a
dominant role in controlling the pressure gradient and the
associated phase structure of the AdS BH system.

\subsection{Specific Heat}
In this part, we shall define the specific heat for the considered
BH configuration. The specific heat is another promising and
thermodynamically worthy measure which assesses the stability of any
BH configuration. Mathematically, the specific heat is defined as
\begin{eqnarray}\nonumber
C=T\left(\frac{\partial S}{\partial
T}\right)=T\left(\frac{dS}{dr_+}\right)\left(\frac{dT}{dr_+}\right)^{-1}
\end{eqnarray}
which, in the present scenario, becomes as follows
\begin{eqnarray}\nonumber
C&=&\frac{r}{{4+32\pi\lambda^2(-1+3r\psi_0)
-2r(6\psi_0-3r\psi_0^2+r\Lambda(2+r\psi_0(-4+3r\psi_0)))}}\left(-2\pi
r(-2+3r\psi_0)
(-2\right.\\\nonumber&+&\left.16\pi\lambda^2+r(4\psi_0-3r\psi_0^2+2r\Lambda(-1+r\psi_0)))+
(-2(\psi_0+24\pi\lambda^2\psi_0)+r(8\Lambda+3\psi_0^2(4-3r\psi_0)\right.\\\nonumber&+&\left.
6r\Lambda\psi_0(-3+2r\psi_0)))\sigma\right).
\end{eqnarray}
\begin{figure}[H]\centering
\includegraphics[width=5.3cm,height=5.0cm]{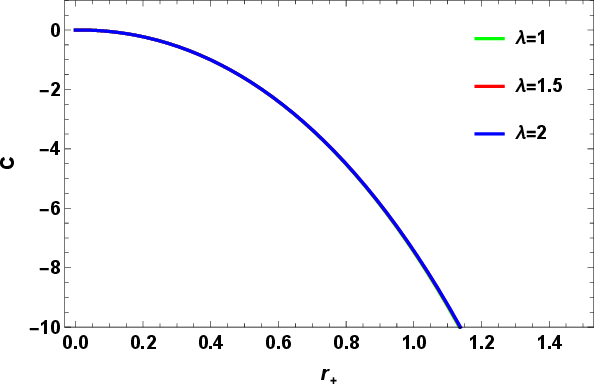}
\includegraphics[width=5.3cm,height=5.0cm]{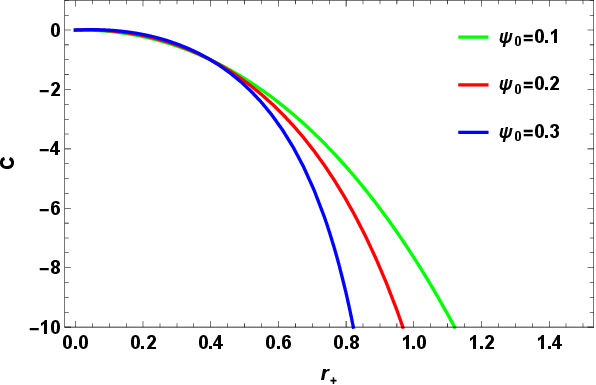}
\includegraphics[width=5.3cm,height=5.0cm]{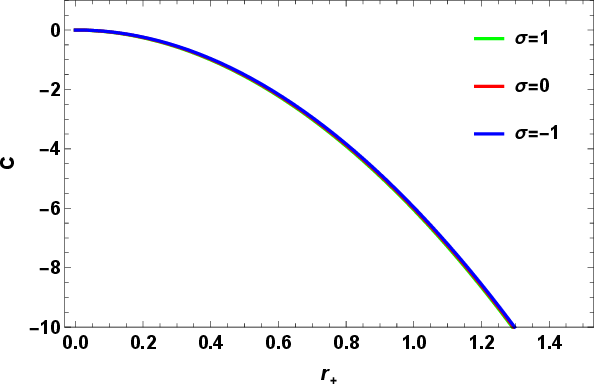}
\caption{Effects of $\lambda$ with fixed $\psi_{0}=0.1$ and
$\sigma=1$ (left panel), $\psi_{0}$ with fixed $\lambda=0.5$ and
$\sigma=1$ (middle panel) and $\sigma$ with fixed $\lambda=0.1$ and
$\psi_{0}=0.05$ (right panel) on specific heat. In all cases, we
fixed $\Lambda=0.005$.}\label{thm8}
\end{figure}
From the plots of Fig.~\textbf{\ref{thm8}}, it is found that
specific heat shows a negative trend versus horizon radius for all
parameter variations, which refer to instability and the BH phase
transition. More specifically, the impact of $\lambda$ and $\sigma$
variations on specific heat are not very obvious, as they produce
similar behavior while for small $\psi_0$, the specific heat is
large and for large $\psi_0$ values, the specific heat is small.

\section{Final Remarks}

Hawking evaporation, shadows and thermodynamic quantities of AdS BHs
are considered as one of the most remarkable subjects of discussion
which have caught the recent interest of numerous researchers. The
present paper is a valuable contribution in this respect, where we
have explored these intriguing phenomena by taking $f(R)$ AdS BH
into account and analyzed our mathematical expressions graphically.
We also graphically investigated the behavior of different important
thermal quantities like corrected entropy, Helmholtz free energy,
internal energy, corrected pressure, enthalpy, Gibbs free energy,
pressure through the Van der Waals model, and specific heat for the
considered AdS $f(R)$ BH. Overall, in the following, we shall summarize
the whole discussion in terms of points.
\begin{enumerate}
\item In the first place, we have computed and analyzed the behavior of the temperature of
AdS $f(R)$ BH versus horizon radius. The respective plots are provided in Fig. \textbf{\ref{evp1}}
which indicated that temperature exhibits different behavior when different values of free parameters
are taken into account. For example, when $\psi_0=0.001,~ \lambda=0.1,~ l=1$, temperature has the maximum
value at $r_{+}=0$ and then suddenly decreases at $r_{+}\approx0.44$ and ultimately sharply increases for
large horizon radius. For negative $\psi_0$, the temperature starts from zero and then behaves linearly with
increasing horizon radius. For large $l$ and positive $\psi_0$ values, the temperature $T$ has a maximum value
at $r_{+}=0$ and then decreases and behaves almost constantly with respect to $r_{+}$. Thus, the selection
of model-free parameters has a noticeable impact on BH temperature.
\item Next, by utilizing the Stefan-Boltzmann law, we have numerically computed the lifetime of BH
evaporation, which turned out to be of order $\sim \ell^3$ approximately. Further, by setting different feasible
choices of model parameters, we have graphically examined the behavior of BH mass against lifetime $t$ to
comprehend the BH evaporation and the respective graphs are provided in Fig. \textbf{\ref{evp2}}. For
$y=0.001$,~$\lambda^{\star}=0.1$, along with $\ell=1$,~$\ell=1.2$ and $\ell=1.4$, it has been noticed
that the BH mass declines rapidly at early times but the evaporation becomes increasingly suppressed
as the BH mass and temperature decrease and consequently, the BH will take infinite time to completely
evaporate away. Further, the BH lifetime turned out to be larger when we assumed: $y=-0.001$,~$\lambda^{\star}=0.2$
with the same variations of the AdS radius. However, for $y=0.001$,~$\lambda^{\star}=0.1$ along with $\ell=2$,~$\ell=3$
and $\ell=4$, the BH can always evaporate away in a finite amount of time, indicating the BH lifetime is
huge for larger AdS.
\item By utilizing the Lagrangian method, we have evaluated null geodesics and discussed their impact on
the photon sphere around the BH. In order to obtain the most
unstable photon orbits, we have explored the physical behavior of
the effective potential $V(r)$ graphically (Fig.
\textbf{\ref{evp3}}). From these plots, it has been noticed that
both $\psi_{0}$ and $\lambda$ parameters have a strong impact on
the effective potential, i.e., as $V(r)$ increases with the large
variations of $\psi_{0}$ as well as $\lambda$. Plots of Fig.
\textbf{\ref{evp4}}) indicated that for large $\psi_{0}$ variations,
the respective circles of BH shadow move towards the central dark
region. Further, increasing monopole parameter $\lambda$ values have
shown an opposite impact on the radius of the BH shadow, i.e., large
$\lambda$ referred to the large BH shadow radius (Fig.
\textbf{\ref{evp5}}). Collectively, it can be concluded that both
parameters are significant in measuring the size of the BH shadow. The
visual signatures of BH shadows are presented in Figs.
{\bf\ref{fal1}} and {\bf\ref{fal2}} for different values of
$\psi_{0}$ and $\lambda$, respectively. The results indicates that
with augmentation $\psi_{0}$, the corresponding shadow radius
decreases, while it is enhanced, with the increase of $\lambda$.

\item Figure \textbf{\ref{thm1}} demonstrates that the graphical behavior of
corrected entropy for AdS BH with respect to the event horizon $r_{+}$
for different values of the involved parameters. From these results, it
is noticed that, for a small domain of $r_{+}$, the entropy exhibits
a negative trend and hence refers to instability as well as phase
transition with the variations of $\lambda$. However, the entropy
exhibits a positive trend for all values of $\psi_0$, while it is
negative only when $\sigma=-1$ at smaller values of $r_{+}$. Figure
\textbf{\ref{thm2}} illustrates the physical behavior of HFE as
a function of the event horizon $r_+$ under different variations of
the model parameters. Here, the impact $\lambda$ and $\psi_{0}$ on
the HFE are qualitatively similar, and depict negative behavior,
which leads to a stable BH thermodynamic configuration. Whereas,
the parameter $\sigma$ introduces a distinct thermodynamic behavior,
particularly in relation to phase transitions at a fixed horizon
radius. Specifically, $\sigma=-1$, the HFE attains positive values
and does not exhibit any phase transition. On the other hand, for
$\sigma=1$, the HFE initially takes negative values, indicating a
stable configuration. Figure \textbf{\ref{thm3}} shows that with the
variations of both $\lambda$ and $\psi_0$, the internal energy turns
out to be a negative quantity for all domains of the event horizon, and thus
it can be concluded that in the present framework, the internal energy
refers to instability. Moreover, the parameter $\psi_0$ produces a
more pronounced separation between the curves than $\lambda$,
suggesting that the gravitational modification plays a more dominant
role in determining the internal energy of the system.

\item Figure \textbf{\ref{thm4}} shows that the
corrected pressure remains positive under the variations of both
$\lambda$ and $\psi_{0}$, while it becomes negative with $\sigma$.
This suggests that the contributions from $\lambda$ and $\psi_{0}$
are more prominent than those from $\sigma$. At $P=0$, the system
exhibits critical behavior at certain values of the horizon radius
$r_+$, indicating a transition between stable and unstable phases.
The obtained results of corrected enthalpy $H$ as presented in
Fig.~\textbf{\ref{thm5}}, indicates that the system exhibits a
negative trend with the variations in both $\lambda$ (only when $r_+
\lesssim 1$), and $\sigma$, while it attains a positive trend in the
case of $\psi_0$. The negative behavior of enthalpy in the small
horizon radius regime suggests that the corresponding BH
configurations are thermodynamically unstable and may correspond to
non-preferred or transitional phases. In contrast, the positive
enthalpy for larger horizon radius indicates physically viable and
stable configurations, which are thermodynamically favored in the
AdS background. Therefore, the system undergoes a transition from an
unstable small BH phase to a stable large BH phase as the horizon
radius increases. The graphical behavior of the GFE is illustrated
in Fig.~\textbf{\ref{thm6}}. The obtained results show a phase
transition between globally stable configurations to an unstable regime
with the variations of $\lambda$. And in the case of $\psi_0$, the
curve exhibits a characteristic U-shaped profile with a distinct
minimum. This behavior signifies that the system is
thermodynamically unstable at both small and large horizon radii.
The variations of $\sigma$, the GFE becomes positive in the small
horizon radius regime when $\sigma =-1$, implying that such
configurations are thermodynamically unstable, whereas stability may
be achieved as the horizon radius increases, and the GFE are
decreases. The expression for change in corrected pressure $\nabla
P$ has been analyzed in the Fig.~\textbf{\ref{thm7}}. The obtained
results demonstrate that variations in $\lambda$ have a more
pronounced effect on $\nabla P$ compared to $\psi_0$ and $\sigma$,
indicating that $\lambda$ plays a dominant role in controlling the
pressure gradient and the associated phase structure of the AdS BH
system. In Fig.~\textbf{\ref{thm8}}, we observed that the specific
heat follows the negative trend with respect to $r_+$, which
indicates that the BH becomes thermodynamically unstable.
\end{enumerate}
In conclusion, the present analysis demonstrates that each model
parameter significantly affects the corresponding Hawking
evaporation, Shadows, and thermodynamic quantities. The results
highlight the importance of incorporating corrected thermodynamics
to investigate microscopic interaction, particularly in the
small and large BH regimes, where quantum and thermal fluctuations
cannot be ignored. In this regard, modified gravity theories provide
a robust avenue for exploring the intricate properties of BHs and
offer deeper insights into both shadow formation and thermodynamic
behavior within the AdS framework, especially in the case of $f(R)$
gravity. Additionally, the analysis of BH shadows suggests that
different BH models can produce distinct observational signatures,
thereby offering valuable information regarding morphology and
dynamics of geometrically thin accretion matter in the surrounding
of celestial objects. An interesting avenue for future research, it
would be interesting to investigate the present analysis by rotating
BHs within the framework of $f(R)$ gravity, as well as to explore
other modified gravity models in order to gain a more comprehensive
understanding of BH dynamics.\\
{\bf Acknowledgements}\\
Princess Nourah bint Abdulrahman University Researchers Supporting
Project number (PNURSP2026R59), Princess Nourah bint Abdulrahman
University, Riyadh, Saudi Arabia.\\
\textbf{Data Availability Statements}\\

No new data has been generated or used in this manuscript.

\end{document}